\documentclass[twocolumn,showpacs,preprintnumbers, amsmath,amssymb,amsfonts, final, a4paper,aps, citeautoscript,footinbib,prb,superscriptaddress]{revtex4-1}
\usepackage[pdftex]{graphicx,color}
\usepackage[latin1]{inputenc}
\usepackage{mathrsfs}
\usepackage{epstopdf}
\usepackage{times}
\usepackage{float}
\usepackage[vcentermath]{youngtab}
\usepackage[sort&compress]{natbib}
\usepackage[colorlinks,bookmarks=true,citecolor=blue,linkcolor=blue,urlcolor=blue, breaklinks=true]{hyperref}
\graphicspath{{}}
\begin{document}
%%%%%%%%%%%%%%%%%%%%%%%%%%%%%%%%%%%%%%%%%%%%%%%%%%%%%%%
\title{Competing superconducting instabilities in the one-dimensional $\boldsymbol{p}$-band degenerate
cold fermionic system}
%%%%%%%%%%%%%%%%%%%%%%%%%%%%%%%%%%%%%%%%%%%%%%%%%%%%%%%
\author{V. Bois}
\affiliation{Laboratoire de Physique Th\'eorique et Mod\'elisation, CNRS UMR 8089,
Universit\'e de Cergy-Pontoise, Site de Saint-Martin,
F-95300 Cergy-Pontoise Cedex, France.}
\author{S.\ Capponi} \affiliation{Laboratoire de Physique Th\'eorique, CNRS UMR 5152, 
Universit\'e Paul Sabatier, F-31062 Toulouse, France.}
\author{P. Lecheminant}
\affiliation{Laboratoire de Physique Th\'eorique et
 Mod\'elisation, CNRS UMR 8089,
Universit\'e de Cergy-Pontoise, Site de Saint-Martin,
F-95300 Cergy-Pontoise Cedex, France.}
\author{M. Moliner}
\affiliation{Universit\'e de Strasbourg, Strasbourg, France.}
%%%%%%%%%%%%%%%%%%%%%%%%%%%%%%%%%%%%%%%%%%%%%%%%%%%%%%%
\date{\today}

% PACS used:
% 71.10.Pm Fermions in reduced dimensions 
% 75.10.Pq Spin chain models
%%%%%%%%%%%%%%%%%%%%%%%%%%%%%%%%%%%%%%%%%%%%%%%%%%%%%%%
\begin{abstract}
The zero-temperature phase diagram of $p$-orbital two-component fermionic system loaded into
a one-dimensional optical lattice is mapped out by means of analytical and numerical techniques.
It is shown that the $p$-band model  away from half-filling 
hosts  various competing superconducting phases for attractive and repulsive interactions.
At quarter filling, we analyze the possible formation of incompressible Mott phases and in particular for repulsive interactions, we find the occurrence of a Mott transition with the formation of fully gapped bond-ordering waves.
\end{abstract}
%%%%%%%%%%%%%%%%%%%%%%%%%%%%%%%%%%%%%%%%%%%%%%%%%%%%%%%
\maketitle
%%%%%%%%%%%%%%%%%%%%%%%%%%%%%%%%%%%%%%%%%%%%%%%%%%%%%%%
\section{Introduction}
%%%%%%%%%%%%%%%%%%%%%%%%%%%%%%%%%%%%%%%%%%%%%%%%%%%%%%%

Ladder systems have been the focus of much theoretical
and experimental work over more than two decades. One theoretical motivation 
was to investigate the dimensional crossover between the well-known 
one-dimensional (1D) physics and the two-dimensional (2D) case, in the search of 2D non-Fermi liquid physics. 
A second reason stems from experiments and the study of ladder compounds, such as the famous telephone number one 
Sr$_{14- x}$Ca$_x$Cu$_{24}$O$_{41}$  which has a superconducting phase at high pression and for a small
hole density. \cite{dagotto}

The simplest ladder model is that of a two-leg ladder,  made of two coupled fermionic
chains. In stark contrast to the single chain case, the two-leg ladder system displays a superconducting
phase with $d$-wave superconductivity for repulsive interactions which stems from the doping of
a spin-gapped Mott insulating phase at half-filling. \cite{bookboso,giamarchi}This gave the belief 
that the two-leg ladder problem already contains seeds of the rich physics of the cuprates. 

Two-leg ladders have thus become over the years a fundamental system for the study of low-dimensional strongly correlated  fermions. Various exotic quantum phases have been predicted theoretically  depending on the form
of the coupling between the two chains and the filling.\cite{larkin,fabrizio,rice,schulz2leg,fisher2leg,orignacdisorder,lin,schulzlast, furusaki,fradkin,lee,rvb1,fabrizio2005,marston,essler,nonnehund,shura}
Experimental realizations of two-leg ladder systems are clearly called for, in particular, to investigate the
rich physics of the weak-coupling regime. In this respect, ultracold fermionic gases are a promising
way to study two-leg ladder problems thanks to the high level of control on interchain hopping and interactions.
\cite{lehur,lewenstein}
The ladder geometry might  be created by considering double-well optical lattices
for instance.\cite{anderlini,danshita,bloch}

\begin{figure}[hbt]
\begin{center}
\includegraphics[scale=0.203]{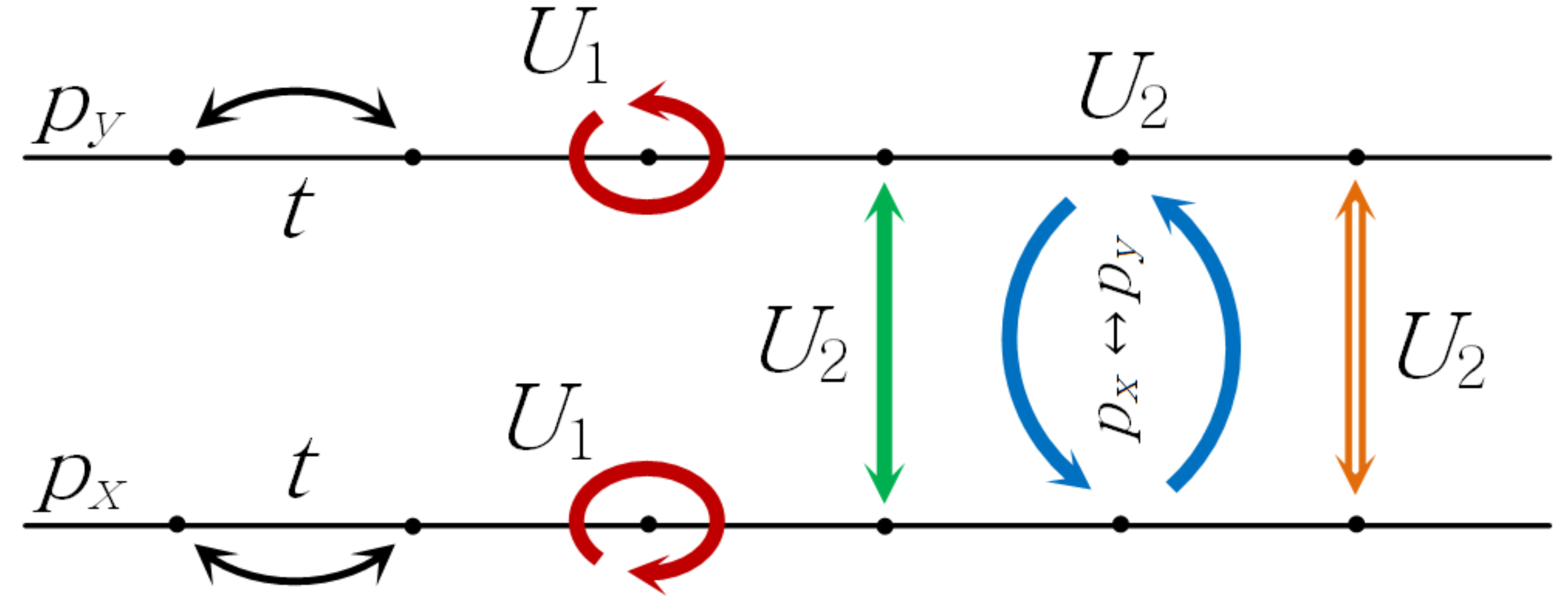}
\caption{(color online) The two-leg fermionic ladder representation of the $p$-band model (\ref{pbandmodel}). 
The orbital degrees of freedom plays the role of the legs of the ladder. Besides the kinetic term with amplitude $t$ for each orbital, there are local interactions at each site $i$: density-density intra-orbital (amplitude $U_1$, in red) and inter-orbital (amplitude $U_2$, in green), orbital exchange (amplitude $U_2$, in blue) and inter-orbital pair-hopping that breaks U(1)$_o$ generically (amplitude $U_2$, in orange).
\label{fig:p-band-2leg}}
\end{center}
\end{figure}

A second possible way is to load a two-component Fermi gas in a optical lattice
and consider higher-lattice orbitals, typically the $p$-band, to simulate  a fermionic two-leg
ladder system. More precisely, we consider, in this paper, a two-component Fermi gas 
which is loaded in a 1D optical lattice (running along the $z$-direction) with moderate strength of (harmonic) confining potential
$m \omega^2 (x^2 +y^2)/2$ in the direction (i.e., $xy$) perpendicular to the chain.  \cite{Kobayashi2012,Kobayashi2014,bois}
It is assumed that all the $s$-level of the oscillator are fully occupied while the $p$-level, i.e., $p_{x,y}$, are partially filled.
The resulting lattice fermionic model has been derived in Refs. \onlinecite{Kobayashi2012,Kobayashi2014,bois} 
within the tight-binding approximation and takes the following form:
\begin{eqnarray}
  H_{p\textrm{-band}} &=& - t\sum_{i,m \alpha} \left(c_{m\alpha,\,i}^\dag c_{m\alpha,\,i+1}+  H.c.\right) \nonumber \\  
&& - ~ \mu\sum_i n_i + \frac{U_1 + U_2}{4} \sum_i  n_i^2 \nonumber \\
&& + ~ \sum_i \left[ 2 U_2 (T_i^x)^2 + (U_1 -U_2) (T_i^z)^2\right] , 
  \label{pbandmodel}
\end{eqnarray}
where $m=p_x, p_y$ is the orbital index and $\alpha =\:\uparrow, \downarrow$ is the
''spin'' index or internal components of the underlying cold atoms. In Eq.~(\ref{pbandmodel}),  $n_i =  \sum_{m\alpha} 
c_{m \alpha,\,i}^\dag c_{m \alpha,\,i}$ describes the density operator at site $i$ 
and a pseudo-spin operator for the orbital degrees of freedom has been defined:
\begin{equation}
  T_i^a =\frac{1}{2}c_{m \alpha,\,i}^\dag \sigma^a_{m n} c_{n \alpha,\,i} ,
  \label{defpeoperator}
\end{equation}
where $\sigma^a, a =x,y,z$ are the Pauli matrices and a summation over repeated indices is implied.
Model  (\ref{pbandmodel}) can be viewed as a two-chain fermionic system without interchain hopping
but with density-density interchain interactions and pair-hopping processes between the two orbitals
(see Fig. \ref{fig:p-band-2leg}).

%%%%%%%%%%%%%%%%%%%%%%%%%%%%%%%%%%%%%%%%%%%%%%%%%%%%%%%%%%
%%%%%%%%%%%%%%%%%%%%%%%%%%%%%%%%%%%%%%%%%%%%%%%%%%%%%%%

The continuous symmetry of  model (\ref{pbandmodel}) for 
general $U_{1,2}$ is: $\text{U}(1)_c\times\text{SU}(2)_s$, where U(1)$_c$ denotes the U(1) (charge) symmetry
related to the conservation of the total number of atoms, while $\text{SU}(2)_s$ is the internal global 
SU(2) (spin) symmetry of the two-component Fermi gas. We note that the U(1)$_{o}$ continuous symmetry
for the orbital degrees of freedom is explicitly broken in Eq.~(\ref{pbandmodel}). 
However, the use of harmonic potential in the $xy$ direction implies a constraint of the two coupling constants $U_{1,2}$ and the investigation of model (\ref{pbandmodel}) along the harmonic line $U_1 = 3 U_2$ enjoys an U(1)$_{o}$ symmetry corresponding to rotation along the $y$-axis in the orbital subspace. The U(1)$_o$  symmetry is trivially realized when $U_2=0$ where the model reduces to two decoupled SU(2) Hubbard chains. Finally, the $U_1 = U_2$ case displays also
an U(1)$_{o}$ symmetry  since it is directly related to $U_2=0$ after a redefinition of the orbital pseudo-spin operator.
We consider, in this paper, the most general case where $U_{1,2}$ are not fine-tuned for two main reasons.
On the one hand, departure from the harmonic line  $U_1 = 3 U_2$ can be  investigated by breaking
the axial symmetry of the 2D harmonic trap with the introduction of quartic potentials.~\cite{bois}
On the other hand,  the harmonic line has been wrongly identified in Refs.~\onlinecite{Kobayashi2012,Kobayashi2014}
and we want to make contact with their results obtained when  $U_2 =4U_1/9$. 

At half-filling, it has been shown that model (\ref{pbandmodel}) along the harmonic
line with repulsive interaction describes a Haldane phase \cite{haldane} of the spin-1 Heisenberg chain. \cite{Kobayashi2012}
The latter is known to be the paradigmatic 1D topological phase protected by symmetry.\cite{oshikawa}
For an attractive interaction, a Haldane phase for the charge degrees of freedom \cite{nonne2010,nonne2011} has also
been predicted in Ref.~\onlinecite{Kobayashi2014} by means of a strong-coupling approach and density-matrix renormalization group (DMRG) calculations.~\cite{white}
In Ref.~\onlinecite{bois}, it has been shown that there is adiabatic continuity between weak and
strong coupling regime and  these two Haldane phases are related by a spin-charge interchange symmetry.
The half-filled $p$-band model (\ref{pbandmodel}) along the harmonic line thus paves the way to realize 
experimentally non-trivial Haldane phases in the context of cold fermions. 

In this paper, by means of a low-energy approach and DMRG calculations, 
we map out the zero-temperature phase diagram of the $p$-band model (\ref{pbandmodel}) 
for incommensurate filling and at quarter filling which best avoids three-body looses. 
In stark contrast to the conclusion of Ref. \onlinecite{Kobayashi2014} for attractive interaction, 
we show that the $p$-band model (\ref{pbandmodel}) away from half-filling does not behave as
an attractive 1D Hubbard chain with the formation of a 2$k_F$ CDW but instead is a representative of 
the physics of two-leg fermionic ladders.  In particular, we find that the hallmark of its
phase diagram is the emergence of various competing superconducting
phases for attractive and repulsive interactions as well as a Mott transition with
the formation of fully gapped bond-ordering waves (BOW).

The rest of the paper is organized as follows. In Sec.~\ref{sec:lowenergy}, we present our low-energy approach based on
a mapping onto Majorana fermions and a one-loop renormalization group (RG) analysis.
A zero-temperature phase diagram for model (\ref{pbandmodel}) is  then deduced in the
weakly-interacting regime for incommensurate filling and at quarter filling with 
Fermi vector $k_F =  \frac{\pi}{4 a_0}$ ($a_0$ being the lattice spacing).
In Sec.~\ref{sec:dmrg}, DMRG calculations are carried out to investigate the intermediate and large-coupling
regimes to fully determine the phase diagram of the $p$-band model.
We present then our concluding remarks in Sec.~\ref{sec:conclusion}.

%%%%%%%%%%%%%%%%%%%%%%%%%%%%%%%%%%%%%%%%%%%%%%%%%%%%%%%
\section{The low-energy approach}\label{sec:lowenergy}
%%%%%%%%%%%%%%%%%%%%%%%%%%%%%%%%%%%%%%%%%%%%%%%%%%%%%%%

In this section, we perform a weak-coupling approach to model (\ref{pbandmodel}) 
away from half-filling. We focus on incommensurate filling and a special commensurate filling where
$k_F =  \frac{\pi}{4 a_0}$. The starting point of this approach is the  introduction of  left-right moving Dirac fermions
from the continuum limit of the non-interacting Hamiltonian (\ref{pbandmodel}) with $U_{1,2} =0$:
\begin{equation}
c_{l\alpha,\,i} \rightarrow \sqrt{a_0} (L_{l\alpha}
e^{-i k_F x} + R_{l\alpha} e^{i k_F x} ),
 \label{contlimitdirac}
\end{equation}
with $x =i a_0$, $l=p_x,p_y$, and $\alpha=\uparrow, \downarrow$.
In stark contrast to the half-filled case of Ref. \onlinecite{bois}, there is a ''spin-charge'' separation which
strongly simplifies the analysis and the full Hamiltonian of the $p$-band model (\ref{pbandmodel}) decomposes into two commuting pieces: 
\begin{equation}
{\cal H} = {\cal H}_c + {\cal H}_s,
\label{spinchargesep}
\end{equation}
with $[{\cal H}_c, {\cal H}_s] = 0$, where  ${\cal H}_c$ governs the physical properties of the charge degrees
of freedom and  ${\cal H}_s$ refers for all remaining non-Abelian degrees of freedom, i.e., spin, orbital and 
spin-orbital degrees of freedom.

\subsection{Charge degrees of freedom}

Let us first consider the simplest part in Eq.~(\ref{spinchargesep}), i.e., the charge degrees of freedom.
For incommensurate filling, on general grounds, the charge part takes the form of a Tomonaga-Luttinger  Hamiltonian:
\cite{bookboso,giamarchi}
\begin{equation}
{\cal H}_c = \frac{v_c}{2} \left[\frac{1}{K_c} 
\left(\partial_x \Phi_c \right)^2 + K_c 
\left(\partial_x \Theta_c \right)^2 \right],
\label{luttbis}
\end{equation}
where $v_c$ and $K_c$ are the Luttinger parameters. In this low-energy approach, the charge excitations are 
described by the bosonic field $\Phi_c$ and its dual field $\Theta_c$.
The explicit form of the Luttinger parameters for the $p$-band model can be extracted from the continuum limit
(\ref{contlimitdirac}) after standard calculations. We find:
\begin{eqnarray}
K_c &=&  \frac{1}{\sqrt{1 + g_c/\pi v_F}}
\nonumber \\
v_c &=& v_F\sqrt{1 + g_c/\pi v_F} ,
\label{Luttingerparameters}
\end{eqnarray}
where $v_F = 2t a_0 \sin (k_F a_0)$ is the Fermi velocity and  $g_c = 2 a_0 (U_1 + U_2)$. 

For incommensurate filling, no umklapp term appears and the charge
degrees of freedom display metallic properties in the
Luttinger liquid universality class. \cite{bookboso,giamarchi}
However, at quarter filling, umklapp processes might be generated leading to 
an additional term in Eq.~(\ref{luttbis}). Such perturbation can be found by means of higher orders of 
the RG calculations or from lattice symmetries.  \cite{OrignacCitro03}
In this respect, the one-step translation symmetry  expresses 
as follows in terms of the charge bosonic field: $T_{a_0}:  \Phi_c  \to \Phi_c + \sqrt{\pi/4}$ .
The charge perturbation with the smallest scaling dimension which is compatible with the translation symmetry
is then: ${\cal V}_c = -g_u \cos(\sqrt{16 \pi} \Phi_c)$. 
The full charge Hamiltonian density for the commensurate filling $k_F =  \frac{\pi}{4 a_0}$ becomes then equivalent
to the quantum sine-Gordon model at $\beta^2 = 16 \pi$:
\begin{equation}
{\cal H}_c = \frac{v_c}{2} \left[\frac{1}{K_c} 
\left(\partial_x \Phi_c \right)^2 + K_c 
\left(\partial_x \Theta_c \right)^2 \right]   -g_u \cos(\sqrt{16 \pi} \Phi_c) .
\label{chargeham}
\end{equation}
This model is exactly solvable and the development of the strong-coupling regime with a charge gap $\Delta_c$ 
corresponds to $K_c  <1/2$, while for $K_c  >1/2$ the charge degrees of freedom remain gapless in
the Luttinger universality class. Using the estimate (\ref{Luttingerparameters}), 
we find the position of the critical line for a Mott transition within the bosonization approach:
\begin{equation}
U_{1c} + U_{2c} \simeq  \frac{3 \pi t \sqrt{2}}{2}  , 
\label{chargeMott}
\end{equation}
such that $K_c =1/2$.  One has to be very careful with this estimate since the full expression of $K_c$ as a function
of $U_{1,2}$  (\ref{Luttingerparameters}) is strictly valid only in the weak-coupling regime. One needs a complementary
approach, i.e., numerical calculations to conclude on the existence or not of a Mott transition for the lattice
model (\ref{pbandmodel}). Note that the mechanism of the Mott transition is very similar
to the one which occurs in the repulsive U(4) Hubbard chain \cite{assaraf99,rey} or in the problem
of spin-3/2 cold fermions. \cite{Lecheminant2005,phle,Sylvain2007,Roux2009}

For $K_c  <1/2$, the charge degrees of freedom are fully gapped and the strong-coupling
regime of the sine-Gordon model (\ref{chargeham}) is described by the pinning of the charged field on the minima:
\begin{eqnarray}
\langle \Phi_{c} \rangle &=& p \sqrt{\pi/4} , \;  g_u > 0
\nonumber \\
\langle \Phi_{c} \rangle &=& \left(p+ \frac{1}{2} \right)  \sqrt{\pi/4} , \;  g_u < 0 ,
\label{pinnedcharge}
\end{eqnarray}
$p$ being an integer. 
The sign of the umklapp coupling constant $g_u$ is difficult to fix but 
we expect $g_u < 0$ from an higher-order perturbative expansion in the weak-coupling regime
so that:$\langle \Phi_{c} \rangle =  \frac{\sqrt{\pi}}{4}$. \cite{OrignacCitro03}
The DMRG calculations of Sec.~\ref{sec:dmrg} will shed light on the position of the pinning of the charge field 
from the determination of the Mott-insulating phases of the $p$-band model. 
The low-lying excitations are massive kinks and antikinks which interpolate between
the ground states of the quantum sine-Gordon model  (\ref{chargeham}). 
The charges associated to these charge excitations are
 \begin{equation}
Q = \pm \frac{2}{\sqrt{\pi}} \int\mathrm{d}x \;  \partial_x \Phi_c =  \pm 1,
\label{chargekinkcharge}
\end{equation}
in units of the electron charge. We have thus massive holon as low-lying excitations in the charge sector.

\subsection{Non-Abelian sector}

The low-energy physics of the remaining degrees of freedom, included in ${\cal H}_{\rm s}$ of 
Eq.~(\ref{spinchargesep}), can be inferred from a mapping onto  six Majorana fermions as
in Refs. \onlinecite{lee,rvb1,fabrizio2005,Azaria99,Boulat}

A simple way to obtain this correspondence is through the introduction
of four chiral bosonic fields $\Phi_{l\sigma R,L}, l= p_x,p_y;\sigma= \: \uparrow, \downarrow$ from 
the bosonization of Dirac fermions:\cite{bookboso,giamarchi}
\begin{eqnarray}
R_{\,l\sigma}&=&\frac{\kappa_{l\sigma}}
{\sqrt{2\pi a_0}}\exp{\left(i\sqrt{4\pi}
\Phi_{l\sigma R}\right)}\nonumber\\
  L_{l\sigma}&=&\frac{\kappa_{l\sigma}}
  {\sqrt{2\pi a_0}}\exp{\left(-i\sqrt{4\pi}
  \Phi_{l\sigma L}\right)},
\label{bosofer}
\end{eqnarray}
where the bosonic fields satisfy the following
commutation
relation:
\begin{equation}
\left[\Phi_{l\sigma R}, \Phi_{l'\sigma' L}\right]
= \frac{i}{4}\delta_{l l'}\delta_{\sigma \sigma'}.
\label{commutator}
\end{equation}
The presence of the Klein factors
$\kappa_{l\sigma}$ ensures the correct
anticommutation of the fermionic operators.
The Klein factors satisfy the anticommutation
rule $\{\kappa_{l\sigma},\kappa_{l'\sigma'}\}
=2\delta_{l l'}\delta_{\sigma \sigma'}$ and they
are constrained so that $\Gamma^2 = 1$, with
$\Gamma=\kappa_{p_x \uparrow}\kappa_{p_x\downarrow}
\kappa_{p_y\uparrow}\kappa_{p_y\downarrow}$.
Hereafter, we will work within the $\Gamma =1$
sector. 

The next step of the approach is to introduce a bosonic basis which singles out
the different degrees of freedom, i.e., charge, spin, orbital, and
spin-orbital degrees of freedom:\cite{Azaria99}
\begin{eqnarray}
  &&\Phi_{p_x\uparrow L,R}=\frac{1}{2}
  \big(\Phi_c+\Phi_s+\Phi_o+\Phi_{so}\big)_{L,R} \nonumber\\
  &&\Phi_{p_x\downarrow L,R}=\frac{1}{2}
  \big(\Phi_c-\Phi_s+\Phi_o-\Phi_{so}\big)_{L,R} \nonumber\\
  &&\Phi_{p_y\uparrow L,R}=\frac{1}{2}
  \big(\Phi_c+\Phi_s-\Phi_o-\Phi_{so}\big)_{L,R} \nonumber\\
  &&\Phi_{p_y\downarrow L,R}=\frac{1}{2}
  \big(\Phi_c-\Phi_s-\Phi_o+\Phi_{so}\big)_{L,R}.
  \label{basebosons}
\end{eqnarray}

From these new bosonic fields, one can now consider a refermionization procedure
by introducing six left and right moving Majorana fermions through:
\begin{eqnarray}
  &&\xi_L^2+i\xi_L^1=
  \frac{\eta_1}{\sqrt{\pi a_0}}
  \exp{\Big(-i\sqrt{4\pi}\Phi_{sL}\Big)}\nonumber\\
  &&\xi_R^2+i\xi_R^1=
  \frac{\eta_1}{\sqrt{\pi a_0}}
  \exp{\Big(i\sqrt{4\pi}\Phi_{sR}\Big)}\nonumber\\
  &&\xi_L^4-i\xi_L^5=
  \frac{\eta_2}{\sqrt{\pi a_0}}
  \exp{\Big(-i\sqrt{4\pi}\Phi_{oL}\Big)}\nonumber\\
  &&\xi_R^4-i\xi_R^5=
  \frac{\eta_2}{\sqrt{\pi a_0}}
  \exp{\Big(i\sqrt{4\pi}\Phi_{oR}\Big)}\nonumber\\
  &&\xi_L^6+i\xi_L^3=
  \frac{\eta_3}{\sqrt{\pi a_0}}
  \exp{\Big(-i\sqrt{4\pi}\Phi_{soL}\Big)}\nonumber\\
  &&\xi_R^6+i\xi_R^3=
  \frac{\eta_3}{\sqrt{\pi a_0}}
  \exp{\Big(i\sqrt{4\pi}\Phi_{soR}\Big)},
   \label{refer}
\end{eqnarray}
where $\eta_{1,2,3,4}$ are again Klein factors
which ensure the adequate anticommutation rules
for the fermions. Using this correspondence, the interacting part of ${\cal H}_{\rm s}$ can
be expressed in terms of these Majorana fermions:
\begin{eqnarray}
&& {\cal H}^{\rm int}_s = 
\frac{\lambda_1}{2}\left(\sum_{a=1}^3\xi^a_R\xi^a_L\right)^2 + \lambda_2\left(\sum_{a=1}^3\xi^a_R\xi^a_L\right)\xi^4_R\xi^4_L \nonumber \\
&& + ~ \lambda_3\left(\sum_{a=1}^3\xi^a_R\xi^a_L\right)\xi^6_R\xi^6_L + \lambda_4\left(\sum_{a=1}^3\xi^a_R\xi^a_L\right)\xi^5_R\xi^5_L \nonumber \\ 
&& + ~ \lambda_5\xi^5_R\xi^5_L\xi^6_R\xi^6_L + \lambda_6\xi^4_R\xi^4_L\xi^5_R\xi^5_L + \lambda_7\xi^4_R\xi^4_L\xi^6_R\xi^6_L , 
 \label{Majomodel}
\end{eqnarray}
with the identification:
\begin{equation}
\begin{array}{lll}
\lambda_1 &=& - a_0(U_1+U_2)\nonumber \\
\lambda_2 &=& -2a_0U_2 \nonumber \\
\lambda_3 &=& a_0(U_2-U_1) \nonumber \\
\lambda_4 &=& 0 \nonumber \\
\lambda_5 &=& 2a_0U_2 \nonumber  \\
\lambda_6 &=& a_0\left(U_1 - U_2\right) \nonumber\\
\lambda_7 &=& 0 .
\label{couplingsN=2}
\end{array}
\end{equation}

Using standard calculations, the one-loop RG equations for model (\ref{Majomodel}) can
de derived:
\begin{eqnarray}
\dot{\lambda}_1 &=& \frac{1}{2\pi}\lambda_1^2 + \frac{1}{2\pi}\lambda_2^2 +\frac{1}{2\pi}\lambda_3^2 + \frac{1}{2\pi}\lambda_4^2  \nonumber \\
\dot{\lambda}_2 &=& \frac{1}{\pi}\lambda_1\lambda_2 + \frac{1}{2\pi}\lambda_3\lambda_7 +\frac{1}{2\pi}\lambda_4\lambda_6 
\nonumber \\
\dot{\lambda}_3 &=& \frac{1}{\pi}\lambda_1\lambda_3 + \frac{1}{2\pi}\lambda_2\lambda_7 +\frac{1}{2\pi}\lambda_4\lambda_5
\nonumber \\
\dot{\lambda}_4 &=& \frac{1}{\pi}\lambda_1\lambda_4 + \frac{1}{2\pi}\lambda_2\lambda_6 +\frac{1}{2\pi}\lambda_3\lambda_5 
\nonumber \\
\dot{\lambda}_5 &=& \frac{3}{2\pi}\lambda_3\lambda_4 + \frac{1}{2\pi}\lambda_6\lambda_7 \nonumber \\
\dot{\lambda}_6 &=& \frac{3}{2\pi}\lambda_2\lambda_4 + \frac{1}{2\pi}\lambda_5\lambda_7  \nonumber \\
\dot{\lambda}_7 &=& \frac{3}{2\pi}\lambda_2\lambda_3 +\frac{1}{2\pi}\lambda_5\lambda_6  .
\label{RGN=2}
\end{eqnarray}

As often in 1D, the RG equations enjoy some hidden discrete symmetries:
 \begin{eqnarray}
 \Omega_1 &:& \lambda_{2,3,4}  \rightarrow - \lambda_{2,3,4} \nonumber  \\
\Omega_2 &:& \lambda_{2,6,7}  \rightarrow - \lambda_{2,6,7}  \nonumber \\
\Omega_3 &:& \lambda_{4,5,6} \rightarrow - \lambda_{4,5,6}  \nonumber \\
\Omega_4 &:& \lambda_{3,5,7} \rightarrow - \lambda_{3,5,7},
\label{eq:symmRG2}
\end{eqnarray}
which correspond to the existence of duality symmetries on the Majorana fermions \cite{Boulat}
\begin{eqnarray}
\Omega_1 &:& \xi^{1,2,3}_{L} \rightarrow -\xi^{1,2,3}_{L} \nonumber  \\
\Omega_2 &:& ~~ \xi^{4}_{L} ~~\, \rightarrow -\xi^{4}_{L} \nonumber  \\
\Omega_3 &:& ~~ \xi^{5}_{L} ~~\, \rightarrow -\xi^{5}_{L} \nonumber  \\
\Omega_4 &:& ~~ \xi^{6}_{L} ~~\, \rightarrow -\xi^{6}_{L} \,\, ,
\label{dualitiespband}
\end{eqnarray}
while the right-moving Majorana fermions remain invariant. These chiral transformations for each Majorana fermion
correspond, in the continuum limit, to the well-known Kramers-Wannier duality symmetry of the underlying 
1D Ising model in a transverse field. The latter transformation maps the ordered phase to the disorder one which
is ordered in terms of the disorder operator. \cite{bookboso} 
Starting from the ordered phase of the six underlying Ising models in a transverse field, the transformations (\ref{dualitiespband}) lead then to four possible different gapful phases. A numerical RG analysis of  Eqs. (\ref{RGN=2})
is necessary to determine which phases is indeed reached starting from the initial conditions (\ref{couplingsN=2}).

\begin{figure}[!ht]
\centering
\includegraphics[width=0.95\columnwidth,clip]{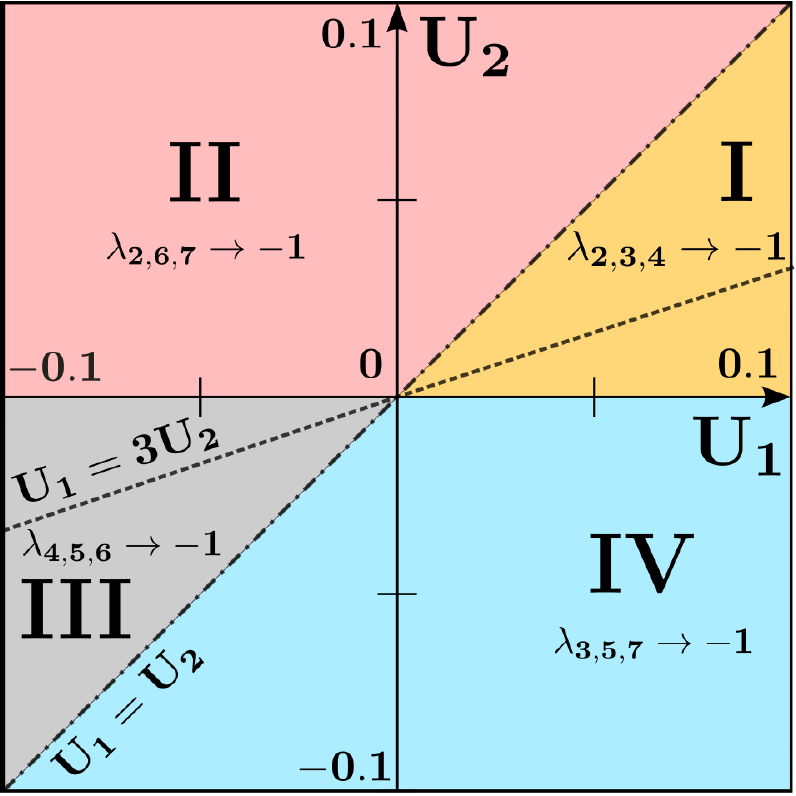}
\vskip 0.3cm
\includegraphics[width=0.95\columnwidth,clip]{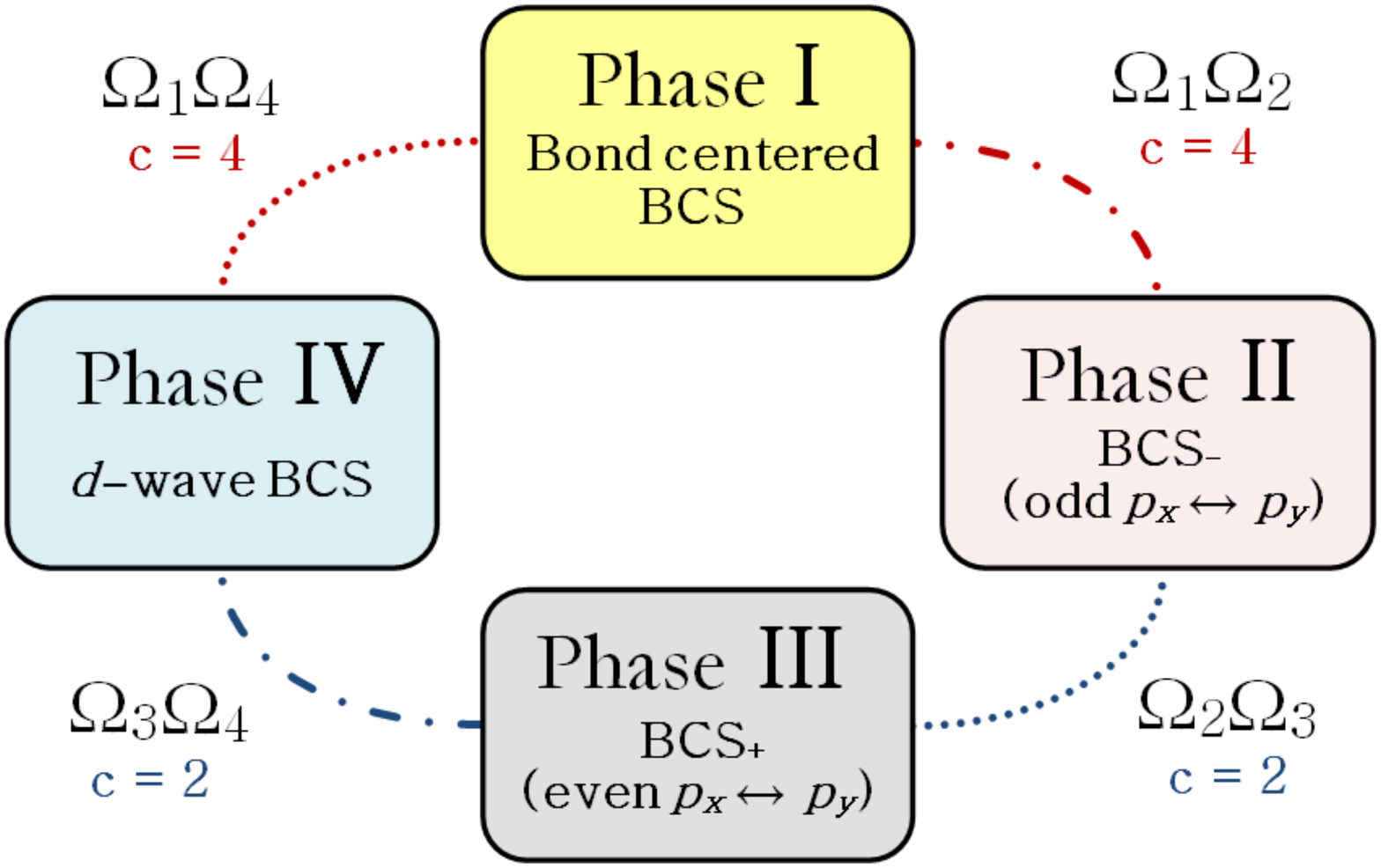}
\caption{(color online) Top: Numerical phase diagram obtained by the one-loop RG analysis of Eq.~(\ref{RGN=2}). Bottom: For each phase denoted I-IV, we give the dominant superconducting correlations (see text) as well as the duality symmetries which
relate the phases between them. We display the central charge $c=2$ or $c=4$ which  characterizes the
different quantum phase transitions of the model.}
\label{fig:Phasediag_N=2}
\end{figure}

We did this numerical analysis and, as depicted in Fig. (\ref{fig:Phasediag_N=2}), it reveals that
the RG flow goes in the strong-coupling regime in the far infrared (IR) along four asymptotic lines:
\begin{eqnarray}
I:\;  && \lambda_2 = \lambda_3 = 
\lambda_4= - \lambda_5 =  - \lambda_6 = - \lambda_7= - \lambda_1 \nonumber \\
II:\;  && \lambda_2 = - \lambda_3 = 
- \lambda_4 = - \lambda_5 =   \lambda_6  =  \lambda_7 = - \lambda_1 \nonumber \\
III:\;  && 
\lambda_2 = \lambda_3 = 
- \lambda_4 = - \lambda_5  =  - \lambda_6  =  \lambda_7 = \lambda_1 \nonumber \\
IV:\;  &&
\lambda_2 = - \lambda_3 = 
 \lambda_4 = - \lambda_5  =  \lambda_6  = - \lambda_7 = \lambda_1 .
\label{asymptotRG}
\end{eqnarray}
Along  these rays, the interacting part of the effective Hamiltonian (\ref{Majomodel}) simplifies as follows:
\begin{eqnarray}
\Omega_1 &:& \mathcal{H}^{\Omega_1}_{\text{int}} = \frac{g}{2}\left(\sum_{a=4}^6\xi^a_R \xi^a_L 
- \sum_{a=1}^3\xi^a_R\xi^a_L\right)^2 \nonumber  \\
\Omega_2 &:& \mathcal{H}^{\Omega_2}_{\text{int}} 
= \frac{g}{2}\left(\sum_{a\neq4}\xi^a_R \xi^a_L - \xi^4_R \xi^4_L \right)^2 \nonumber  \\
\Omega_3 &:& \mathcal{H}^{\Omega_3}_{\text{int}} 
= \frac{g}{2}\left(\sum_{a\neq5}\xi^a_R \xi^a_L - 
\xi^5_R \xi^5_L \right)^2 \nonumber  \\
\Omega_4 &:& \mathcal{H}^{\Omega_4}_{\text{int}} 
= \frac{g}{2}\left(\sum_{a\neq6}\xi^a_R \xi^a_L - \xi^6_R \xi^6_L \right)^2 ,
\label{SO6raysapp}
\end{eqnarray}
with $g>0$. 
Using the duality symmetries (\ref{dualitiespband}), all these models reduces to the same model which takes the
form of the SO(6) Gross-Neveu (GN) model with interaction: \cite{GN}
\begin{eqnarray}
{\cal H}^{\rm int}_{\rm GN} &=& 
\frac{g}{2} \left(  \sum_{a=1}^{6} \xi^{a}_R \xi^{a}_L \right)^2 .
   \label{GNSO6}
\end{eqnarray}
This is an example of a dynamical symmetry enlargement (DSE) where an
SO(6) symmetry, i.e. a higher symmetry than the initial symmetry of the $p$-band lattice model,  emerges at low-energy.  
This phenomenon occurs in a large variety of models with marginal interactions 
in the scaling limit \cite{saleur} as, for instance,
in the half-filled two-leg Hubbard model \cite{lin}
and in the SU(4) Hubbard chain at half-filling \cite{assaraf}, where an SO(8) symmetry occurs at low energy.
The emergence of an SO(6) symmetry has also been obtained within the one-loop RG analysis of 
various models of doped two-leg ladders. \cite{schulzlast,lee,rvb1,essler,Boulat}

\subsection{Phases of the $p$-band model}

The main interest of this SO(6) DSE scenario stems from the fact that the 
isotropic RG ray is described by a massive integrable field theory (\ref{GNSO6}) when $g >0$.
\cite{zamolo,karowski} 
The development of the strong-coupling regime in the SO(6) model leads to generation of a non-perturbative fermionic mass,
i.e. the emergence of a spin-gap phase where all spin, orbital and spin-orbital excitations are fully gapped. 
The nature of the underlying electronic phases of the $p$-band model can then be inferred 
by a straightforward semiclassical approach of the SO(6) model and the application of the duality 
symmetries (\ref{dualitiespband}). Alternatively, one can perform a direct bosonic semiclassical approach of the different models in Eq.~(\ref{SO6raysapp}) using the identification (\ref{refer}).  In the following, we proceed to this analysis by identifying the phases of Fig.~\ref{fig:Phasediag_N=2}.

\subsubsection{Phase I: $\Omega_1$ duality}

In the first phase, when $U_1 > 0$ and $U_1 > U_2$ of Fig. (\ref{fig:Phasediag_N=2}),  
the numerical analysis of the one-loop  RG equations reveals that the RG flow is attracted in the far 
IR along a special ray  (I) of Eq. (\ref{asymptotRG}).
This phase is described by the $\Omega_1$ duality and the resulting physical properties of that phase
are governed by the interacting Hamiltonian $\mathcal{H}^{\Omega_1}_{\text{int}} $ of Eq.~(\ref{SO6raysapp}).
 A straightforward semiclassical approach of the latter model  leads to the following pinning of the bosonic fields:
 \begin{equation}
 \langle \Phi_s \rangle =  0, \;   \; \langle \Phi_o \rangle = \frac{\sqrt{\pi}}{2},   \;  \; \langle \Theta_{so} \rangle =  0 .
 \label{pinningI}
 \end{equation}
 The electronic properties of phase I, which includes the harmonic line $U_1 = 3U_2$, 
 depend also on the charge degrees of freedom that are decoupled.
 
 Let us first consider the $K_c > 1/2$ case.
One has then a gapless charge excitation whereas all remaining spin, orbital, and spin-orbital excitations are fully gapped.
 All $2k_F$ densities are short-ranged due to the pinning of the dual spin-orbital field $\Theta_{so}$. Interestingly enough, the leading electronic instability in this phase turns out to be a bond-centered superconducting instability with order parameter:
 \begin{equation}
{\cal O}_{\tiny{\mbox{BCSbond}}} = c_{p_x\uparrow,i+1}c_{p_y\downarrow,i}
-c_{p_x\downarrow,i+1}c_{p_y\uparrow,i}
- (p_x \leftrightarrow p_y) ,
\label{BCSs}
\end{equation}
which is odd under the $\mathbb{Z}_2$ orbital exchange symmetry ($p_x \leftrightarrow p_y$).
A similar superconducting instability has been introduced in the study of doped two-leg electronic ladder in Ref.~\onlinecite{Robinson2012}.
In the continuum limit, we have:
\begin{eqnarray}
{\cal O}_{\tiny{\mbox{BCSbond}}} &\simeq& a_0 2i \sin(k_F a_0) \big( 
R_{p_x\uparrow} L_{p_y\downarrow} + R_{p_y\downarrow} L_{p_x\uparrow}
\\
&& ~~~~~~~~~~~~~~~~~~~~~~~ - ~ R_{p_x\downarrow} L_{p_y\uparrow}  - R_{p_y\uparrow} L_{p_x\downarrow} \big). \nonumber
\label{BCSscont}
\end{eqnarray}
One can then bosonize this operator by means of the identification (\ref{bosofer}) 
and the physical basis (\ref{basebosons}):
\begin{eqnarray}
{\cal O}_{\tiny{\mbox{BCSbond}}} &\simeq&   \frac{4 \sin(k_F a_0)}{\pi} \kappa_{p_x\uparrow} \kappa_{p_y\downarrow} 
e^{- i \sqrt{\pi} \Theta_c} \\
&& \times\big[ - i \sin \sqrt{\pi} \Phi_s \cos \sqrt{\pi} \Phi_o \sin \sqrt{\pi} \Theta_{so} \nonumber \\
&& ~~~~~ + ~ \cos \sqrt{\pi} \Phi_s \sin \sqrt{\pi} \Phi_o \cos \sqrt{\pi} \Theta_{so} \big]. \nonumber
\label{BCSsboson}
\end{eqnarray}
Taking account of the vacuum expectation values (\ref{pinningI}), we get
${\cal O}_{\tiny{\mbox{BCSbond}}} \simeq   e^{- i \sqrt{\pi} \Theta_c} $ so that the equal-time correlation
function of the pairing operator has a power-law decay:
 \begin{equation}
 \langle {\cal O}^{\dagger}_{\tiny{\mbox{BCSbond}}} \left(x\right)  {\cal O}_{\tiny{\mbox{BCSbond}}} \left(0\right) \rangle
 \sim x^{-1/2K_c} .
\label{BCSscorrel}
\end{equation}
When $1/2 < K_c < 1$, the bond-centered BCS superconducting
instability is strongly enhanced with respect to the non-interacting case. \emph{We have thus a dominant superconducting instability for repulsive interaction}.

While the $2k_F$ densities are short-ranged in phase I,
$4k_F$ densities might compete with the superconducting instability (\ref{BCSs}).
Many $4k_F$ density terms can be written in the continuum description
like $ \rho^{(1)}_{4 k_F} \sim  \sum_{l\sigma} 
L^{\dagger}_{l \sigma} R_{l \sigma} L^{\dagger}_{l -\sigma} R_{l -\sigma}$ for instance.
The latter gives in the bosonized language:
\begin{equation}
\rho^{(1)}_{4 k_F} \sim 
e^{ i \sqrt{4\pi} \Phi_c}   \cos \sqrt{4\pi} \Phi_o,
\label{rho4kF1}
\end{equation}
so that in the spin-gapped phase (\ref{pinningI}), one has $\rho^{(1)}_{4 k_F} \sim  e^{ i \sqrt{4\pi} \Phi_c}$, and therefore
a power-law decay for its correlation function:
\begin{equation}
 \langle \rho^{(1) \dagger}_{4 k_F}\left(x\right)  \rho^{(1)}_{4 k_F}  \left(0\right) \rangle
 \sim x^{- 2 K_c} .
\label{4kfcorrel}
\end{equation}
Since  $ K_c >1/2$, one observes that the $4k_F$ density correlation function (\ref{4kfcorrel})
decays much faster than the superconducting one (\ref{BCSscorrel}).
In summary,  the leading instability of phase I with $K_c > 1/2$ is 
the bond-centered BCS superconducting one.

In the Mott-insulating phase with $K_c < 1/2$, all excitations are fully gapped.
The bond-centered BCS superconducting instability (\ref{BCSs})  is now short-ranged  since 
the charge field $\Phi_c$ is pinned (\ref{pinnedcharge}) and thus its dual field
$\Theta_c$ is a strongly fluctuating field.
In contrast,  the $4k_F$ density of  Eq.~(\ref{rho4kF1}) can have now a non-zero expectation
value, breaking spontaneously the translation symmetry.
We thus expect a Mott phase with a two-fold degenerate ground state at quarter filling $k_F =  \frac{\pi}{4a_0}$.
The nature of this phase depends on the sign of the coupling constant  $g_u$ of the
umklapp perturbation in model (\ref{chargeham}).  In this respect, we introduce the following order parameters
as in the study of the extended quarter-filled two-leg Hubbard ladder: \cite{OrignacCitro03}
\begin{eqnarray}
 {\cal O}_{\rm BOW_{\pi,q_y}}\!(i) &=& \left(-1\right)^{i} \sum_{\sigma} \Big( c^{\dagger}_{p_x \sigma,i+1} c_{p_x \sigma,i}
 \nonumber  \\
&& ~~~~~~~~~~~~~~ + \cos q_y \; c^{\dagger}_{p_y \sigma,i+1} c_{p_y\sigma,i} + H.c. \Big) \nonumber \\
  {\cal O}_{\rm CDW_{\pi,q_y}}\!(i) &=& \left(-1\right)^{i} \sum_{\sigma} \Big( c^{\dagger}_{p_x \sigma,i} c_{p_x \sigma,i} 
   \nonumber \\
&& ~~~~~~~~~~~~~~ + \cos q_y \; c^{\dagger}_{p_y \sigma,i} c_{p_y\sigma,i}  \Big) , 
 \label{orderparaMottN=2}
\end{eqnarray}
with $q_y = 0, \pi$.
One can obtain a continuum and bosonized representation for these order parameters. The $4k_F$ contribution of the latter
plays a crucial role and can be derived using the results of Refs. \onlinecite{Haldane4kf,Schulz93,OrignacCitro03} and
we get for the leading contribution:
\begin{eqnarray}
&& (-1)^{i} \sum_{\sigma} c^{\dagger}_{l\sigma,i+1} c_{l \sigma,i} +H.c.  \sim  
\sin \sqrt{4\pi} (\Phi_{c} + \epsilon_l \Phi_{o})
\nonumber \\
&& (-1)^{i} \sum_{\sigma} c^{\dagger}_{l\sigma,i} c_{l \sigma,i}   \sim
 \cos \sqrt{4\pi} (\Phi_{c} + \epsilon_l \Phi_{o}) ,
 \label{orbitalkboson}
\end{eqnarray}
with $\epsilon_{p_x} = 1$, $\epsilon_{p_y} = -1$.
From these results, we find the bosonized descriptions of the order parameters (\ref{orderparaMottN=2}):
\begin{eqnarray}
 {\cal O}_{\rm BOW_{\pi,\pi}} &\sim& \cos\sqrt{4 \pi} \Phi_c \sin \sqrt{4 \pi} \Phi_o \nonumber \\
 {\cal O}_{\rm BOW_{\pi,0}} &\sim& \sin\sqrt{4 \pi} \Phi_c \cos\sqrt{4 \pi} \Phi_o \nonumber \\
  {\cal O}_{\rm CDW_{\pi,\pi}} &\sim& \sin\sqrt{4 \pi} \Phi_c \sin \sqrt{4 \pi} \Phi_o \nonumber \\
  {\cal O}_{\rm CDW_{\pi,0}} &\sim& \cos\sqrt{4 \pi} \Phi_c \cos \sqrt{4 \pi} \Phi_o  .
 \label{orderparaN=2boson}
\end{eqnarray}
Taking into account that in phase I, we have $\langle \Phi_o \rangle = \sqrt{\pi}/2$,
we have either a uniform  $4k_F$ BOW or uniform  $4k_F$ CDW depending on the sign of the umklapp term in Eq.~(\ref{chargeham}):
\begin{eqnarray}
 \langle {\cal O}_{\rm BOW_{\pi,0}}   \rangle &\ne& 0 , \;\; {\rm if}  \;\;  g_u < 0 \nonumber \\
 \langle {\cal O}_{\rm{CDW}_{\pi, 0}}  \rangle &\ne& 0 , \;\; {\rm if}  \;\;  g_u > 0 .
  \label{Mottphase1}
\end{eqnarray}
We expect in the weak-coupling regime that $g_u<0$. A  uniform $4k_F$ BOW is thus stabilized which is two-fold degenerate and breaks the translation symmetry. The latter phase is similar to the BOW phase obtained in the 
quarter-filled spin-3/2 SO(5) chain model. \cite{Sylvain2007}

\subsubsection{Phase II: $\Omega_2$ duality}

In the second phase, when $U_2 > 0$ and $U_1 < U_2$,  the one-loop  
RG flow is now attracted in the far IR by the asymptote (II) of Eq. (\ref{asymptotRG}).
This phase is described by the $\Omega_2$ duality. A straightforward semiclassical approach of 
model $\mathcal{H}^{\Omega_2}_{\text{int}} $ in Eq.~(\ref{SO6raysapp}) leads to the following pinning of
the bosonic fields in this phase:
\begin{equation}
 \langle \Phi_s \rangle =  0, \;   \; \langle \Theta_o \rangle = 0,   \;  \; \langle \Phi_{so} \rangle =  0 .
 \label{pinningII}
 \end{equation}
 In this phase, as we will see in Sec.~\ref{sec:dmrg}, numerical results find that  $K_c > 1/2$.
 All spin, orbital, and spin-orbital excitations are fully gapped in this phase and the charge degrees of freedom
 are gapless since $K_c > 1/2$.  Due to the presence
 of attractive interactions, one may call this gapless phase a Luther-Emery phase as in the spin-1/2 Hubbard chain with
 $U<0$. \cite{bookboso,giamarchi}
Its physical nature is very different from that of phase I since  the expectation values of the 
bosonic fields (\ref{pinningII}) are different.
As it can readily be seen from Eq.~(\ref{pinningII}),  we have  again the condensation of a dual field, here the 
orbital one $ \Theta_o$, which implies that the $2k_F$-CDW operator is a strongly fluctuating order. As in phase I, 
the only possible CDW quasi-long range order in this phase is a $4k_F$ CDW. 
However, the $\rho^{(1)}_{4 k_F} $ CDW  of Eq.~(\ref{rho4kF1}) becomes now short-range since
the orbital dual field $ \Theta_o $ condenses in phase II. In this respect, we have to consider another 
$4k_F$ CDW instability:
$\rho^{(2)}_{4 k_F} \sim   \sum_{\sigma} 
L^{\dagger}_{p_x \sigma} R_{p_x \sigma} L^{\dagger}_{p_y \sigma} R_{p_y \sigma}$.
The latter can be directly expressed in terms of the bosonic field:
\begin{equation}
\rho^{(2)}_{4 k_F} \sim  e^{ i \sqrt{4\pi} \Phi_c}   \cos \sqrt{4\pi} \Phi_s. 
\label{rho4kF2}
\end{equation}
Using the expectation values (\ref{pinningII}),
we deduce the leading asymptotics of the equal-time correlation function of the $4k_F$ CDW operator:
\begin{equation}
 \langle \rho^{(2) \dagger}_{4 k_F}\left(x\right)  \rho^{(2)}_{4 k_F}  \left(0\right) \rangle
 \sim x^{- 2 K_c} .
\label{4kfcorrelbis}
\end{equation}
As it will be seen in the next section, we have $K_c >1/2$ and, one does not expect, on general ground, that
this $4k_F$-CDW will be the dominant instability in this phase.  As in phase I, a superconducting instability 
turns out to be the leading one. The latter is defined by the following order parameter:
\begin{equation}
{\cal O}_{\rm BCS-} = c_{p_x\uparrow,i}c_{p_x\downarrow,i}
- c_{p_y\uparrow,i}c_{p_y\downarrow,i},
\label{BCS-}
\end{equation}
which is odd under the $\mathbb{Z}_2$ orbital symmetry ($p_x \leftrightarrow p_y$) and antisymmetric
with respect to the spin degrees of freedom ($\uparrow \leftrightarrow \downarrow$).
The order parameter (\ref{BCS-}) can be expressed directly in terms of the bosonic fields:
\begin{eqnarray}
{\cal O}_{\rm BCS-} &\simeq& \frac{2 \kappa_{p_x \uparrow} \kappa_{p_x \downarrow} }{\pi} 
e^{- i \sqrt{\pi} \Theta_c} \\
&& \times\Big[ \cos \sqrt{\pi} \Phi_s \cos \sqrt{\pi} \Theta_o \cos \sqrt{\pi} \Phi_{so} \nonumber \\
&& ~ +  i \sin \sqrt{\pi} \Phi_s \sin \sqrt{\pi} \Theta_o \sin \sqrt{\pi} \Phi_{so} \Big]. \nonumber
\label{BCS-boson}
\end{eqnarray}
Using the expectation values (\ref{pinningII}), we get
${\cal O}_{\rm BCS-} \simeq   e^{- i \sqrt{\pi} \Theta_c} $, so that the equal-time correlation
function of the pairing operator (\ref{BCS-}) has a power-law decay:
 \begin{equation}
 \langle {\cal O}^{\dagger}_{\rm BCS-} \left(x\right)  {\cal O}_{\rm BCS-} \left(0\right) \rangle
 \sim x^{-1/2K_c} ,
\label{BCS-correl}
\end{equation}
which dominates  the $4k_F$ CDW ones (\ref{4kfcorrelbis}) when $K_c > 1/2$.

\subsubsection{Phase III: $\Omega_3$ duality}

The next phase, defined by $U_2 < 0$ and $U_1 < U_2$,  is described by  the  asymptote (III) of the one-loop RG flow. 
The harmonic line of the $p$-band model with attractive interaction belongs to this phase which is described
by the $\Omega_3$ duality with interacting Hamiltonian  $\mathcal{H}^{\Omega_3}_{\text{int}} $ of Eq.~(\ref{SO6raysapp}).  The bosonic fields of the bosonization approach are now pinned to the values:
 \begin{equation}
 \langle \Phi_s \rangle =  0, \;   \; \langle \Theta_o \rangle = \frac{\sqrt{\pi}}{2},   \;  \; \langle \Phi_{so} \rangle =  0 .
 \label{pinningIII}
 \end{equation}
 A spin-gap is formed and a gapless $c=1$ phase emerges in this attractive regime with $U_{1,2} < 0$ 
 since the umklapp term cannot gap out the charge degrees of freedom when $K_c >1$.  In close parallel to the previous cases, one can determine the nature of the leading electronic instability of this Luther-Emery
phase  by means of the bosonization approach combined with the pinning (\ref{pinningIII}).
 The dominant $4k_F$ CDW is the one of phase II, given by Eq.~(\ref{rho4kF2}) with the power-law behavior 
 (\ref{4kfcorrelbis}).  The relevant superconducting instability for phase III is defined by:
 \begin{equation}
{\cal O}_{\rm BCS+} = c_{p_x\uparrow,i}c_{p_x\downarrow,i}
+ c_{p_y\uparrow,i}c_{p_y\downarrow,i},
\label{BCS+}
\end{equation}
which is even under the $\mathbb{Z}_2$ orbital symmetry ($p_x \leftrightarrow p_y$) and antisymmetric
with respect to the spin degrees of freedom ($\uparrow \leftrightarrow \downarrow$). In terms of the bosonic fields,
it reads as follows:
\begin{eqnarray}
{\cal O}_{\rm BCS+} &\simeq& - \frac{2 \kappa_{p_x\uparrow} \kappa_{p_x\downarrow} }{\pi} e^{- i \sqrt{\pi} \Theta_c} \\
&& \times \Big[i\cos \sqrt{\pi} \Phi_s \sin \sqrt{\pi} \Theta_o \cos \sqrt{\pi} \Phi_{so} \nonumber \\
&& ~~ + \phantom{i}\sin \sqrt{\pi} \Phi_s \cos \sqrt{\pi} \Theta_o \sin \sqrt{\pi} \Phi_{so} \Big]. \nonumber
\label{BCStboson}
\end{eqnarray}
Using the vacuum expectation values (\ref{pinningIII}), we immediately get
${\cal O}_{\rm BCS+} \simeq   e^{- i \sqrt{\pi} \Theta_c} $, so that the equal-time correlation
function of this pairing operator has a power-law decay:
 \begin{equation}
 \langle {\cal O}^{\dagger}_{\rm BCS+} \left(x\right)  {\cal O}_{\rm BCS+} \left(0\right) \rangle
 \sim x^{-1/2K_c} .
\label{BCS+correl}
\end{equation}
Since $K_c >1$, we conclude that the superconducting instability (\ref{BCS+}) dominates
the $4k_F$-CDW ordering (\ref{4kfcorrelbis}). \emph{The  Luther-Emery phase is thus governed
by a superconducting instability (\ref{BCS+}) in stark contrast to the $2k_F$-CDW phase predicted
by the DMRG study of Ref.~\onlinecite{Kobayashi2014}}. The physics of the $p$-band model with attractive
interaction is thus not similar to that of 1D attractive Hubbard model as emphasized in Ref.~\onlinecite{Kobayashi2014}.

\subsubsection{Phase IV: $\Omega_4$ duality}

The last phase of the $p$-band model corresponds to the region where $U_2 < 0$ and $U_1 > U_2$.
The numerical analysis of the RG flow shows that, here, the one-loop  
RG flow is attracted in the far IR by the special line (IV) of Eq. (\ref{asymptotRG}).
The resulting phase is described by the $\Omega_4$ duality with interacting Hamiltonian  $\mathcal{H}^{\Omega_4}_{\text{int}} $ of Eq.~(\ref{SO6raysapp}) which leads to the following pinning for the bosonic fields of 
 the bosonization approach:
 \begin{equation}
 \langle \Phi_s \rangle =  0, \;   \; \langle \Phi_o \rangle = 0,   \;  \; \langle \Theta_{so} \rangle =  0 .
 \label{pinningIV}
 \end{equation}
 As before, this pinning leads to the formation of a gapless $c=1$ phase when $K_c > 1/2$ where
 the charge degrees are the only critical modes of the problem.
 We now consider the standard $d$-wave superconducting instability of the two-leg electronic ladder to
 determine the nature of phase IV:
\begin{equation}
{\cal O}_{\rm BCSd} = c_{p_x\uparrow,i}c_{p_y\downarrow,i}
-c_{p_x\downarrow,i}c_{p_y\uparrow,i},
\label{BCSd}
\end{equation}
which is even under the $\mathbb{Z}_2$ orbital symmetry ($p_x \leftrightarrow p_y$). The bosonized expression of
this  superconducting instability reads as follows:
\begin{eqnarray}
{\cal O}_{\rm BCSd} &\simeq& \frac{2 \kappa_{p_x\uparrow} \kappa_{p_y\downarrow} }{\pi} e^{- i \sqrt{\pi} \Theta_c} \\
&& \times\Big[ i \sin \sqrt{\pi} \Phi_s \sin \sqrt{\pi} \Phi_o \sin \sqrt{\pi} \Theta_{so} \nonumber \\
&& + \cos \sqrt{\pi} \Phi_s \cos \sqrt{\pi} \Phi_o \cos \sqrt{\pi} \Theta_{so} \Big]. \nonumber
\label{BCSdboson}
\end{eqnarray}
From this expression, we observe that this operator is a fluctuating order, i.e., has short-ranged correlation,
in the previous phases, while, in phase IV, one has from the pinning (\ref{pinningIV}):
${\cal O}_{\rm BCSd} \sim e^{- i \sqrt{\pi} \Theta_c} $.
The $d$-wave superconducting instability (\ref{BCSd}) becomes dominant in phase IV with the power-law behavior:
 \begin{equation}
 \langle {\cal O}^{\dagger}_{\rm BCSd} \left(x\right)  {\cal O}_{\rm BCSd} \left(0\right) \rangle
 \sim x^{-1/2K_c} .
\label{BCSdcorrel}
\end{equation}
From the pinning (\ref{pinningIV}), we find that the $2k_F$-CDW operator is short-ranged while the $4k_F$ CDW of phase II, given by Eq.~(\ref{rho4kF2}), has a power-law behavior in phase IV:
\begin{equation}
 \langle \rho^{(2) \dagger}_{4 k_F}\left(x\right)  \rho^{(2)}_{4 k_F}  \left(0\right) \rangle
 \sim x^{- 2 K_c} ,
\label{4kfcorrelfin}
\end{equation}
with subleading exponent when $K_c > 1/2$ with respect to the superconducting instability (\ref{BCSdcorrel}).

\subsubsection{Quantum phase transitions}\label{subsec:qpt}

From the duality symmetries (\ref{dualitiespband}), we can, as well, discuss  the different quantum
phase transitions that occur in the $p$-band model by investigating self-dual manifolds. 
However,  Fig. \ref{fig:Phasediag_N=2} reveals that the transitions belong to special lines of the lattice
model: $U_1= U_2$ and $U_2=0$.  
From Eq.~(\ref{pbandmodel}), the $U_2=0$ describes two decoupled quarter-filled spin-1/2 Hubbard chains  
with $U_1$ coupling constant. When $U_1 <0$, a spin-gap is formed and therefore the
phase II/phase III transition is critical with $c=1+1=2$ gapless charge modes.
When $U_1 > 0$, all degrees of freedom are gapless since the quarter-filled spin-1/2 Hubbard chains  
is known not to exhibit a Mott transition. \cite{giamarchi}
The phase I/phase IV transition is thus critical with central charge $c=4$.
The transition between phase I and phase II is  located along the $U_1 =U_2>0$ line of the $p$-band model.
From the definition of the $p$-band model (\ref{pbandmodel}), one can show that the latter line corresponds to two decoupled 
repulsive quarter-filled spin-1/2 Hubbard chains which does not have relevant umklapp process. 
A $c=4$ behavior should occur along this quantum phase transition.
Finally, the last transition between phase III and phase IV belongs to the $U_2 = U_1< 0$ line which takes the form of
two decoupled attractive quarter-filled spin-1/2 Hubbard chains with a  $c=2$ quantum critical behavior.
All these results can be derived by investigating self-dual manifolds of the RG Eqs.~(\ref{RGN=2}).
Since the transition lines are located on special high symmetry lines, one would expect the location of these transitions to remain universal beyond the weak coupling regime where this analysis is valid. This will be indeed confirmed numerically in Sec.~\ref{subsec:qpt_num}.  A summary of the phases and quantum phase transitions, obtained from
the low-energy approach, can be found in the bottom of Fig. \ref{fig:Phasediag_N=2}.

%%%%%%%%%%%%%%%%%%%%%%%%%%%%%%%%%%%%%%%%%%%%%%%%%%%%%%%
\section{DMRG calculations}\label{sec:dmrg}
%%%%%%%%%%%%%%%%%%%%%%%%%%%%%%%%%%%%%%%%%%%%%%%%%%%%%%%

\subsection{Determination of the phase diagram}
We will now determine the phase diagram of the $N=2$ $p$-band model at quarter filling (i.e. 1 particle per site) using numerical simulations with the  DMRG algorithm. This will allow to go beyond the one-loop RG analysis done in Fig.~\ref{fig:Phasediag_N=2} and check the analytical predictions for realistic intermediate or strong couplings $(U_1,U_2)$ (we fix $t=1$ as the unit of energy).  
Moreover, numerical data are needed to determine the numerical value of the Luttinger parameter $K_c$ which allows to compute the dominant correlations. 

Typically, we have used open boundary
conditions (OBC) and lengths $L=64$ and $L=128$, keeping up to 4000 states when computing correlations in order to keep a discarded weight below $10^{-10}$ in most regions, although simulations where $ 0 \leq U_2 \leq U_1$ were found to be more difficult to converge (discarded weight around $10^{-7}$, see discussion below). 
For practical purpose, we have mapped the $p$-band model onto an  \emph{equivalent} (pseudo)spin-1/2 (where the
pseudo spin corresponds to the orbital) fermionic models on a $2$-leg ladder, and we have implemented the abelian U(1) symmetry corresponding to the conservation of particles spin.

\begin{figure}[!ht]
\centering
\includegraphics[width=\columnwidth,clip]{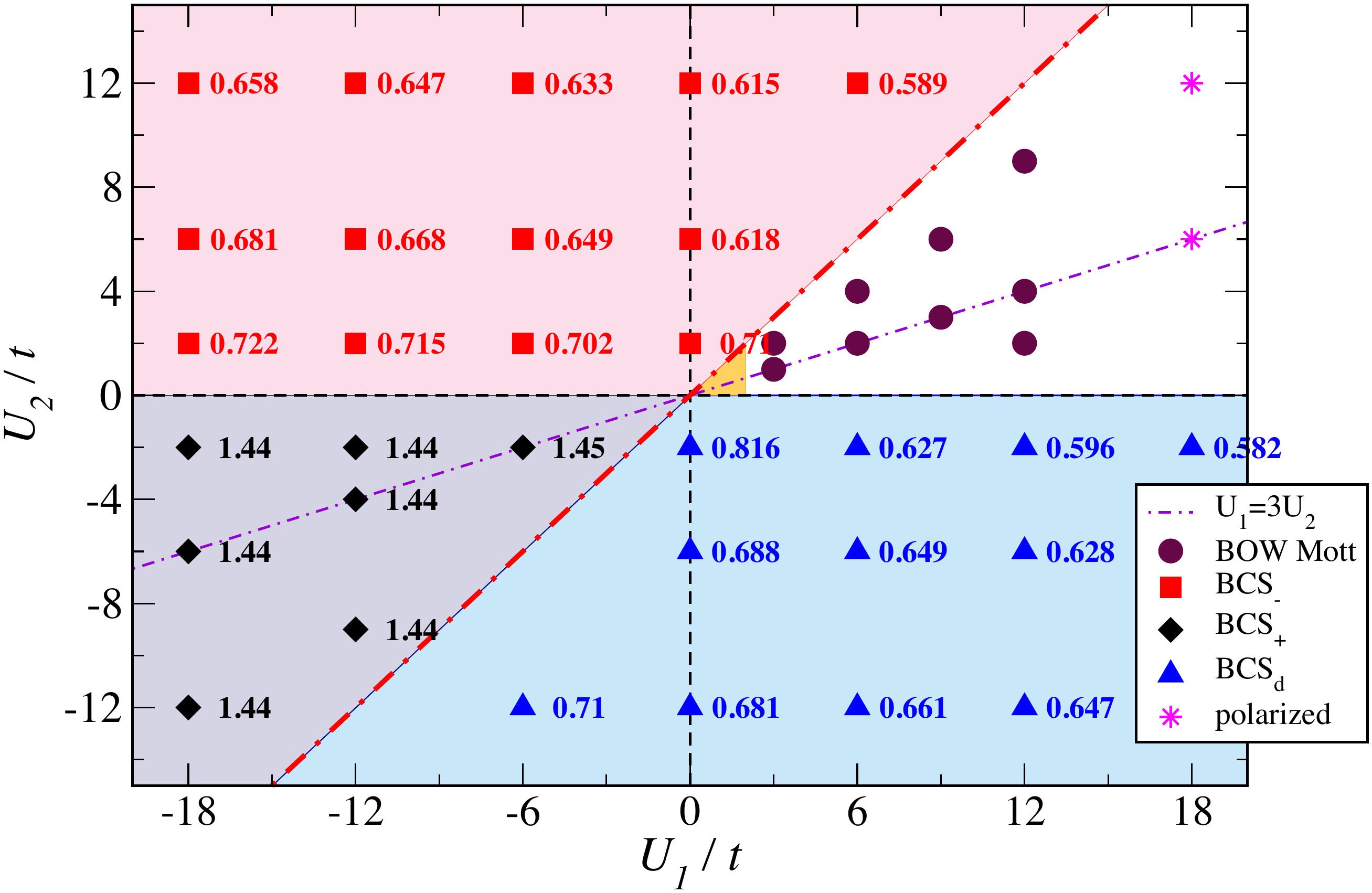}
\caption{(color online) Numerical phase diagram obtained by DMRG simulations with length $L=128$. Numerical values correspond to the Luttinger parameter $K_c$ in each region.}
\label{fig:phasediag_pband_N2_n1}
\end{figure}

Our main result is presented in Fig.~\ref{fig:phasediag_pband_N2_n1} where we plot the phase diagram vs $(U_1/t, U_2/t)$ obtained by computing various correlation functions on $L=128$ systems. We will present the numerical data below, but we can already discuss the different phases. 
First of all, the (quantum phase) transition lines are found to be $U_2=0$ and $U_1=U_2$, which is expected by symmetry since they correspond to special lines of the model, see Sec.~\ref{subsec:qpt}. A detailed analysis will be given in Sec.~\ref{subsec:qpt_num}. Second, it is remarkable that phases II, III and IV found in the weak-coupling analysis are confirmed to exist in a wide range of parameters. Last, we have computed the Luttinger parameter $K_c$ using the dominant correlation, which is always of some BCS type, in each of these phases. 

Overall, we observe a rather good agreement between the numerical phase diagrams obtained by solving RG equations (Fig.~\ref{fig:Phasediag_N=2}) or the full microscopic model using DMRG (Fig.~\ref{fig:phasediag_pband_N2_n1}), except for the large Mott phase at this commensurate filling. For incommensurate filling, we predict that the RG phase diagram would be identical to the numerical one.  

We will now present some numerical data that were used to compute this numerical phase diagram. We have relied mostly on computing superconducting correlation functions (in various channels) as well as density ones. To avoid spurious effects due to OBC, we have chosen to compute correlations as 
\begin{equation}
C_{\cal O}(x) = \langle {\cal O}^\dagger(L/4+x) {\cal O}(L/4) \rangle
\end{equation}
so that we can determine if they decay algebraically or are short-ranged. In Figs.~\ref{fig:corr_U1_m6_U2_6}-\ref{fig:corr_U1_m6_U2_m2}-\ref{fig:corr_U1_6_U2_m6}, such correlations are plotted and correspond respectively to phases II, III and IV. In all cases, we emphasize that dominant correlations are found to be the superconducting ones, in different channels (see below). Indeed, the Luttinger parameter $K_c$ is always larger than 1/2 in the critical phases, otherwise a Mott phase is stabilized. Some values of $K_c$ are given on the phase diagram in Fig.~\ref{fig:phasediag_pband_N2_n1}. In particular, for phases II, III and IV, we observe an adiabatic continuity from weak to strong coupling, while phase I on the contrary is much reduced and replaced instead by Mott phase at intermediate coupling and polarization at strong coupling. 

\subsubsection{Phase II}

\begin{figure}[!ht]
\centering
\includegraphics[width=\columnwidth,clip]{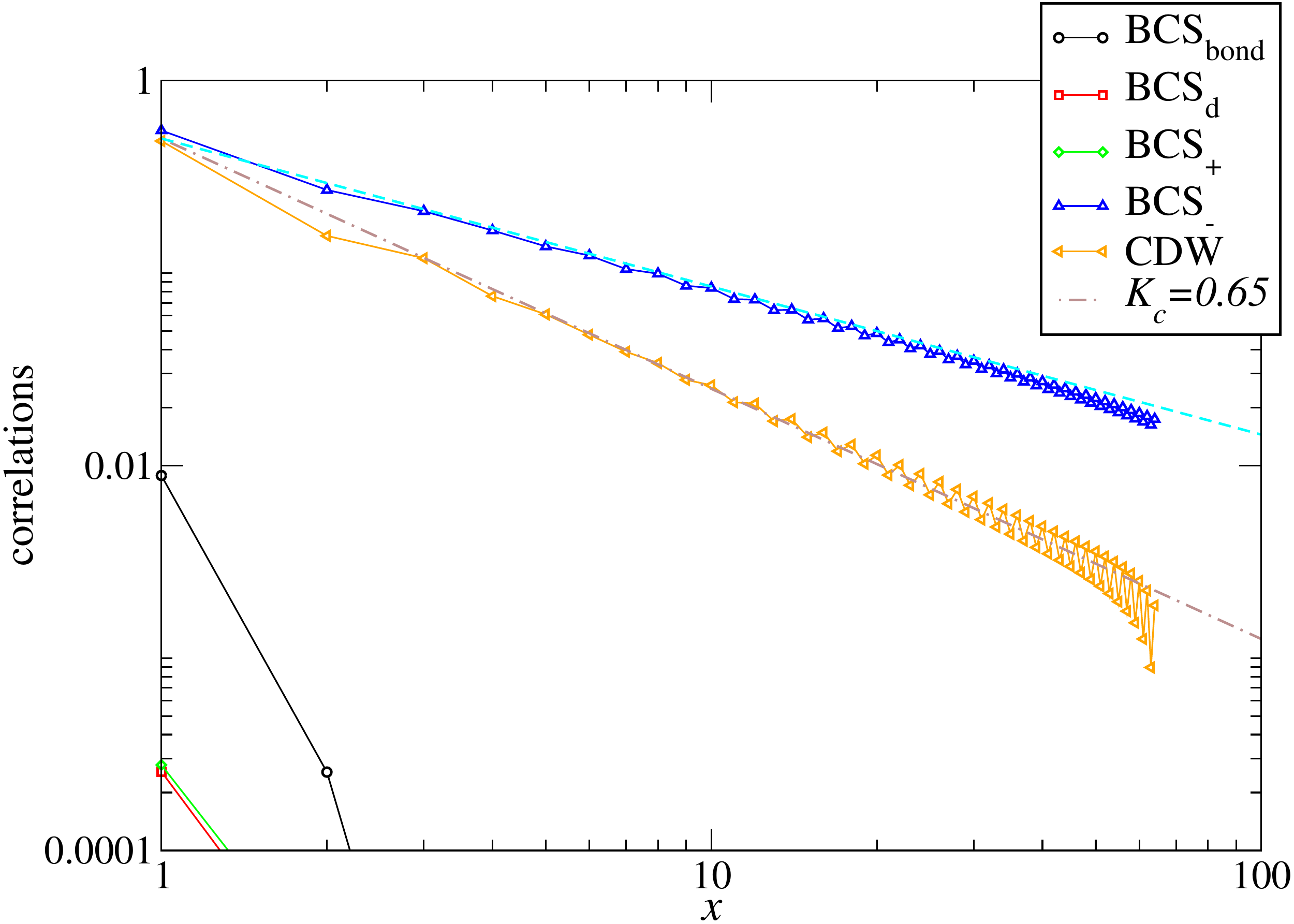}
\caption{(color online) Absolute values of correlation functions (see text) vs distance for ($U_1/t=-6$, $U_2/t=6$) and $L=128$ obtained with DMRG. This corresponds to our phase II. Fitting the dominant BCS$_-$ correlation with Eq.~(\ref{BCS-correl}) leads to $K_c=0.65$, which is also consistent with the $4k_F$ CDW correlation function (see Eq.~(\ref{4kfcorrel})).}
\label{fig:corr_U1_m6_U2_6}
\end{figure}

In this region of the phase diagram, our numerical results, shown in Fig.~\ref{fig:corr_U1_m6_U2_6} for instance when $U_1/t=-6$ and $U_2/t=6$, are in perfect agreement with the RG predictions and dominant BCS$_-$ predictions. However, numerics is needed to compute the Luttinger parameter $K_c$ and determine whether pairing correlations dominate over density ones. Using Eqs.~(\ref{4kfcorrelbis})-(\ref{BCS-correl}), our numerical fits for both BCS$_-$ and CDW correlations are compatible with a single $K_c>1/2$, for instance 0.65 for the parameters set chosen on the plot. All the other correlations are short-ranged as expected. The leading instability in this phase is therefore the BCS$_-$ superconducting one.

\subsubsection{Phase III}

\begin{figure}[!ht]
\centering
\includegraphics[width=\columnwidth,clip]{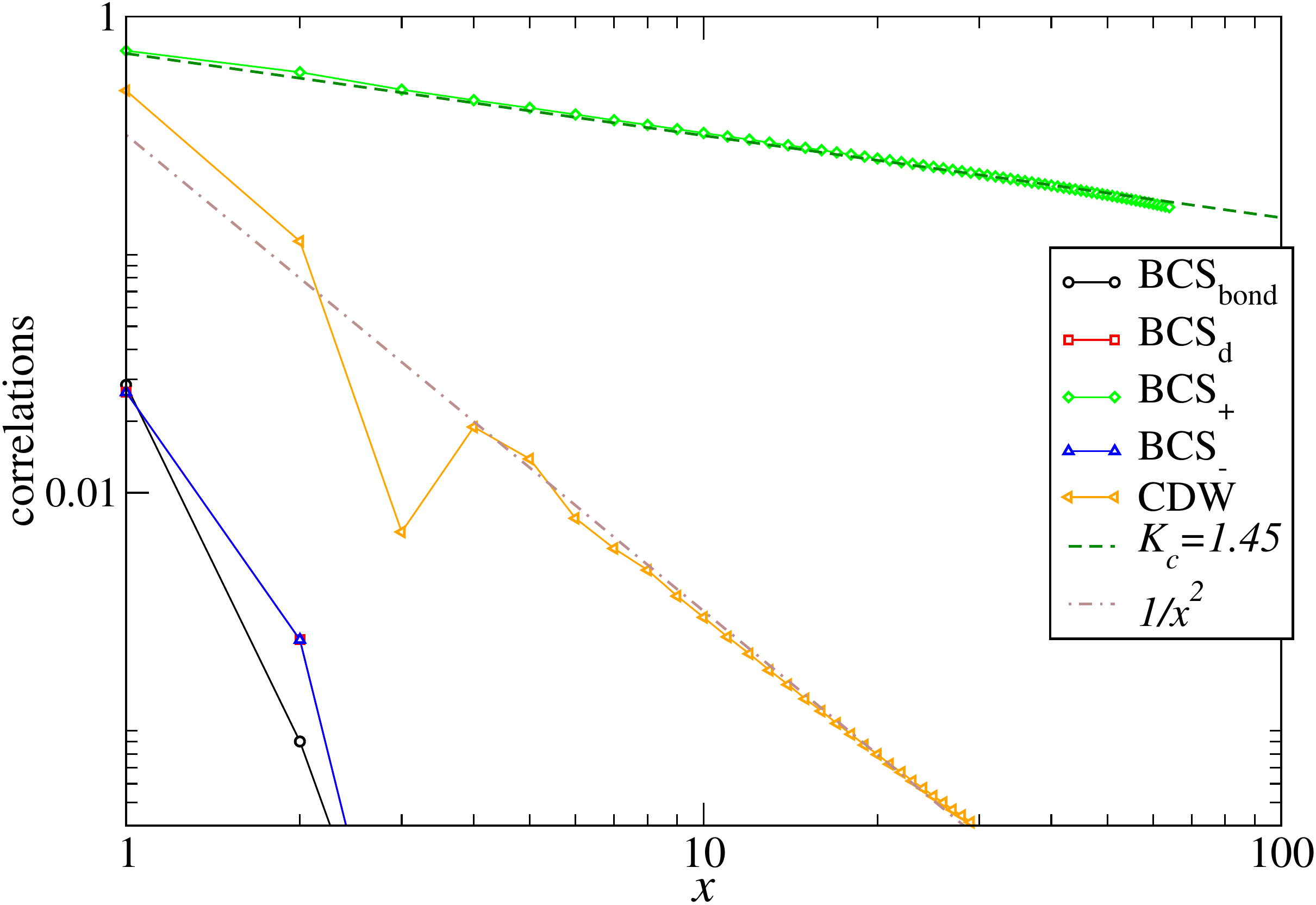}
\caption{(color online) Absolute values of correlation functions (see text) vs distance along the harmonic line in the attractive case ($U_1/t=-6$, $U_2/t=-2$) for $L=128$ obtained with DMRG. This corresponds to our phase III. Fitting the dominant BCS$_+$ correlation with Eq.~(\ref{BCS+correl}) leads to $K_c=1.45$.}
\label{fig:corr_U1_m6_U2_m2}
\end{figure}

The phase III region is particularly interesting since it could be achieved experimentally using the harmonic trapping scheme (i.e. $U_1=3U_2$) with attractive interactions. As seen in Fig.~\ref{fig:corr_U1_m6_U2_m2}, we have found that the dominant correlations are of the pairing type, in the BCS$_+$ channel. In all this region, they can be fitted using Eq.~(\ref{BCS+correl}) to get $K_c>1$ (values are given in the phase diagram). In such a case, the dominant density correlations are \emph{uniform} and decay as $1/x^2$ as expected. 

Note that this dominant superconducting correlation function was not computed in Ref.~\onlinecite{Kobayashi2014}, where only CDW signal was discussed. As a result, the dominant instability in this Luther-Emery phase is a superconducting one, in stark contrast to the  $2k_F$-CDW phase predicted earlier~\cite{Kobayashi2014}.

\subsubsection{Phase IV}

\begin{figure}[!ht]
\centering
\includegraphics[width=\columnwidth,clip]{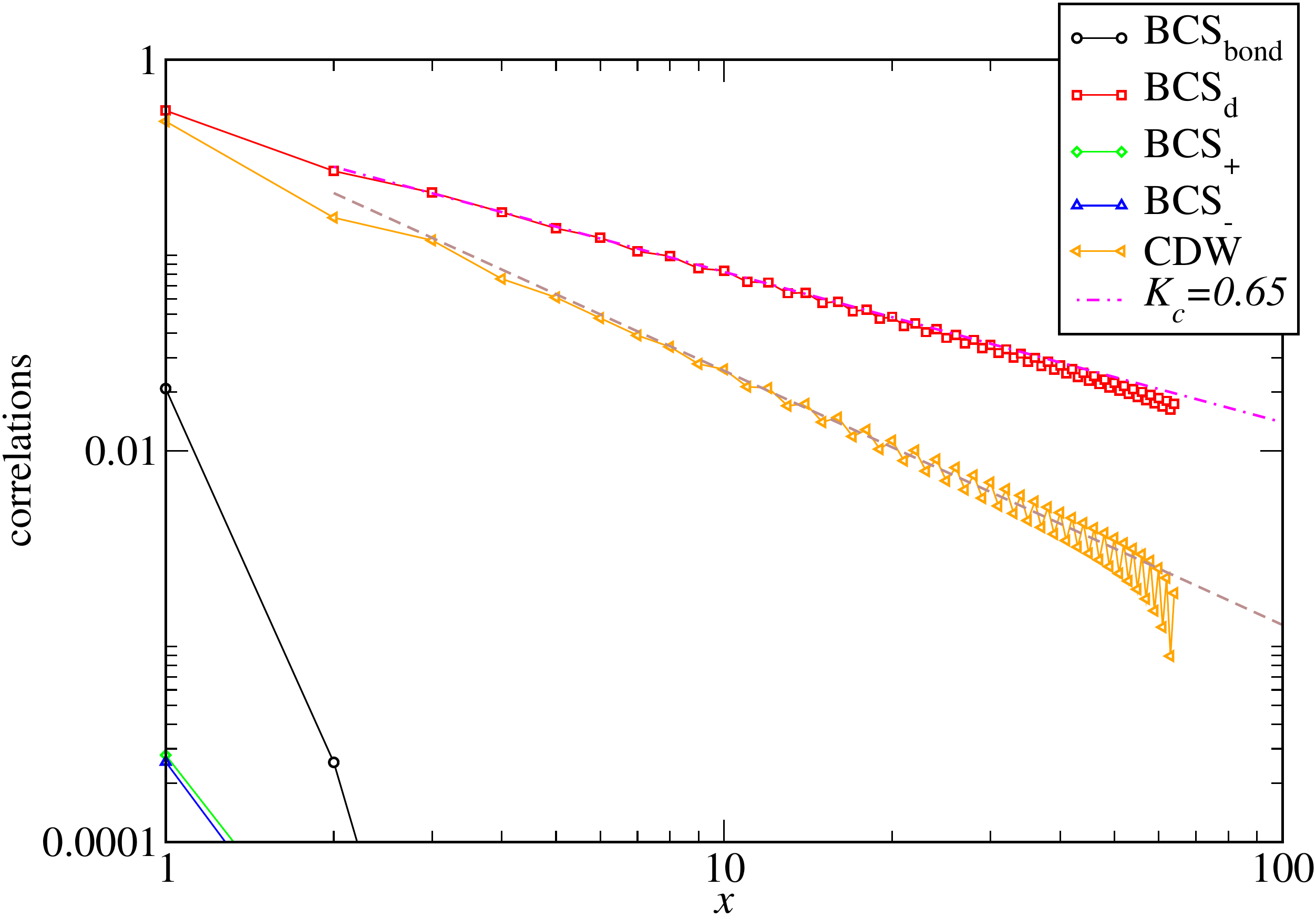}
\caption{(color online) Absolute values of correlation functions (see text) vs distance for ($U_1/t=6$, $U_2/t=-6$) and $L=128$ obtained with DMRG. This corresponds to our phase IV. Fitting the dominant BCS$_d$ correlation with Eq.~(\ref{BCSdcorrel}) leads to $K_c=0.65$, which is also consistent with the $4k_F$ CDW correlation function (see Eq.~(\ref{4kfcorrel})).}
\label{fig:corr_U1_6_U2_m6}
\end{figure}

In the region corresponding to phase IV, our numerical results in Fig.~\ref{fig:corr_U1_6_U2_m6} indicate that dominant correlations are of the pairing type again, but in a different BCS$_d$ channel, as expected from the low-energy analysis. In the whole region, we can fit these correlations using Eq.~(\ref{BCSdcorrel}) to extract $K_c$, or equivalently we could fit the subleading CDW correlations with Eq.~(\ref{4kfcorrelfin}), although it would be less precise. Some values of $K_c$ are given in Fig.~\ref{fig:phasediag_pband_N2_n1} and are always larger than 1/2 so that no Mott transition is found. 

\subsubsection{Phase I}

\begin{figure}[!ht]
\centering
\includegraphics[width=\columnwidth,clip]{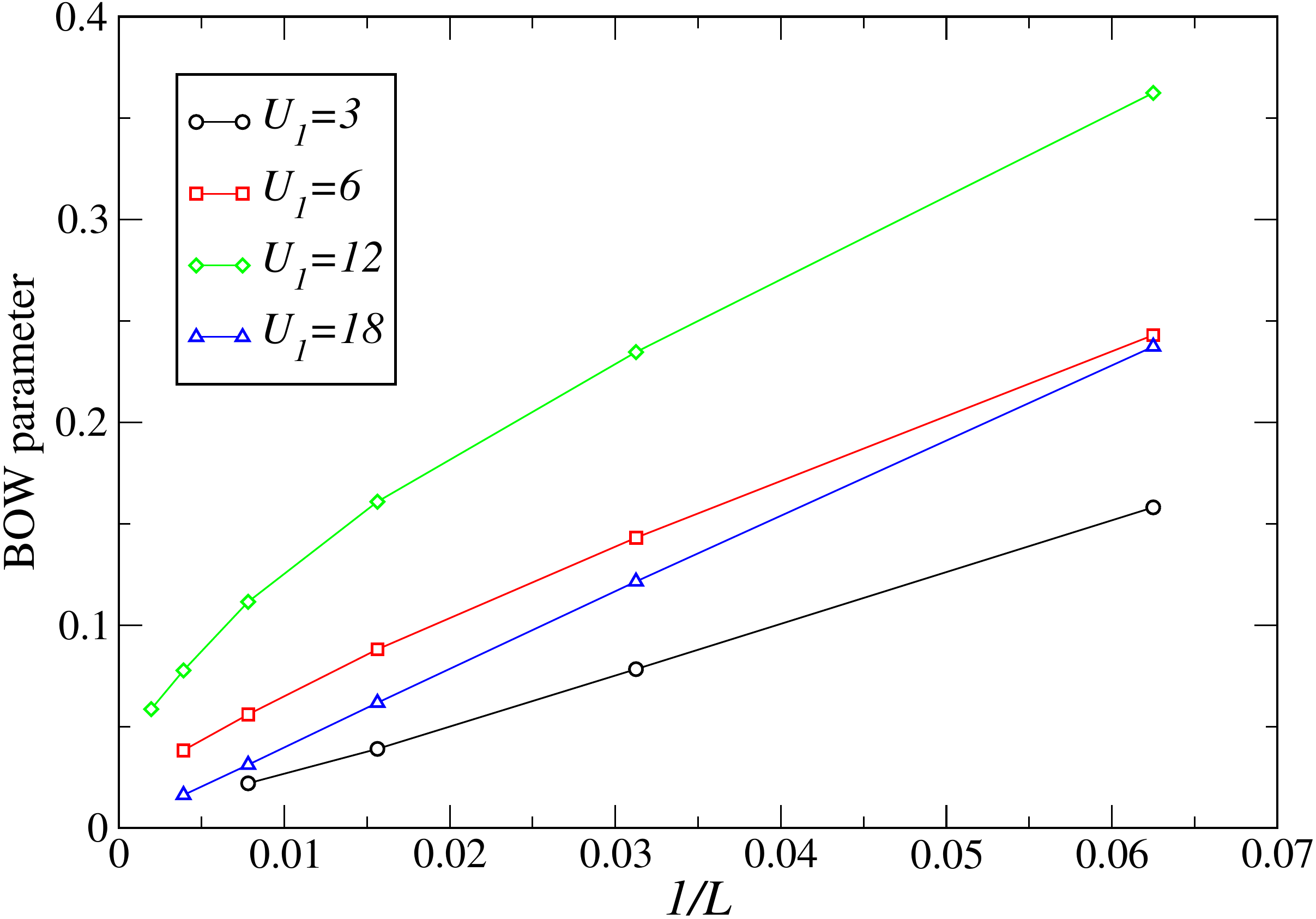}
\caption{(color online) Finite-size scaling of the $(\pi,0)$ BOW order parameter along the harmonic line $U_1=3U_2$.}
\label{fig:bow}
\end{figure}

The region $0<U_1<U_2$, which should correspond to phase I (see Fig.~\ref{fig:Phasediag_N=2}) according to RG solution is more involved to analyze. 
We have plotted on Fig.~\ref{fig:bow} the finite-size scaling of the $(\pi,0)$ BOW, i.e. the kinetic energy difference measured in the middle of the chain. Data are given along the harmonic line $U_1=3U_2$. Extrapolations are compatible with a finite, albeit small, value for intermediate interactions (for instance $U_1=6$, $U_2=2$), while it seems to vanish in the weak and strong coupling regimes. It is known that the weak coupling is difficult numerically since the non-interacting starting point correspond to two decoupled fermionic chains with a large total central charge $c=4$. For instance, the Mott transition in the SU($N$) Hubbard model at filling $1/N$ has been discussed quite extensively to occur for a \emph{finite} critical $(U/t)_c$ when $N>2$ based on bosonization and quantum Monte-Carlo results~\cite{assaraf99} as well as DMRG ones~\cite{rey}, while some older DMRG simulations had indicated a \emph{vanishing} $U_c=0$~\cite{Buchta2007}. Since the charge gap is expected to open in an exponentially way, this is clearly a difficulty for any numerical technique. However, this regime is perfectly suited for bosonization and weak-coupling RG: indeed, since the Luttinger parameter $K_c=1$ in the non-interacting case, umklapp processes are irrelevant so that finite interactions are necessary to enter the Mott phase. 

Thus, our interpretation is the following: (i) for small interaction parameters, RG analysis should be valid and phase I with dominant BCS$_{\mathrm{bond}}$ correlations is expected with $K_c>1/2$ so that umklapp processes are irrelevant; (ii) for intermediate interactions, a Mott phase occurs with $(\pi,0)$ BOW and exponentially decaying BCS correlations at large distance; (iii) for very large interactions (i.e. $U_1=18$ and $U_2=6$), we have noticed that the ground-state is ferromagnetically polarized (hence degenerate): the polarized ground-state can be simply understood as two decoupled spinless fermionic chains (one for each orbital), which energy is independent of both $U_1$ and $U_2$. Intuitively, such a state could be stabilized at large $U$'s since other competing states will have higher energies. 
Still, our numerics is clear on this point as can be checked by comparing ground-state energies in different sectors (varying the number of particles per spin), or measuring $K_c=1$ from charge correlations (data not shown). 

\subsection{Nature of the phase transitions}\label{subsec:qpt_num}

A simple way to obtain information on the phases, or on their phase transitions, is through the measurements of block von Neumann entanglement entropy which is known to scale as~\cite{Calabrese-C-04}:
\begin{equation}\label{eq:Svn}
S_{\text{vN}}=(c/6) \log d(x|L) + \mathrm{Cst}
\end{equation}
where $c$ is the central charge and $d(x|L)=(L/\pi) \sin (\pi x/L)$ the conformal distance. 
On a finite system with OBC, due to Friedel oscillations, the fitting can be more involved and one can use for instance the knowledge of the local kinetic bond energies to get more reliable results (see Fig.~11 of Ref.~\onlinecite{Roux2009} for instance). Our numerical results should be compared to the analytical predictions made in Sec.~\ref{subsec:qpt}.

\begin{figure}[!ht]
\centering
\includegraphics[width=\columnwidth,clip]{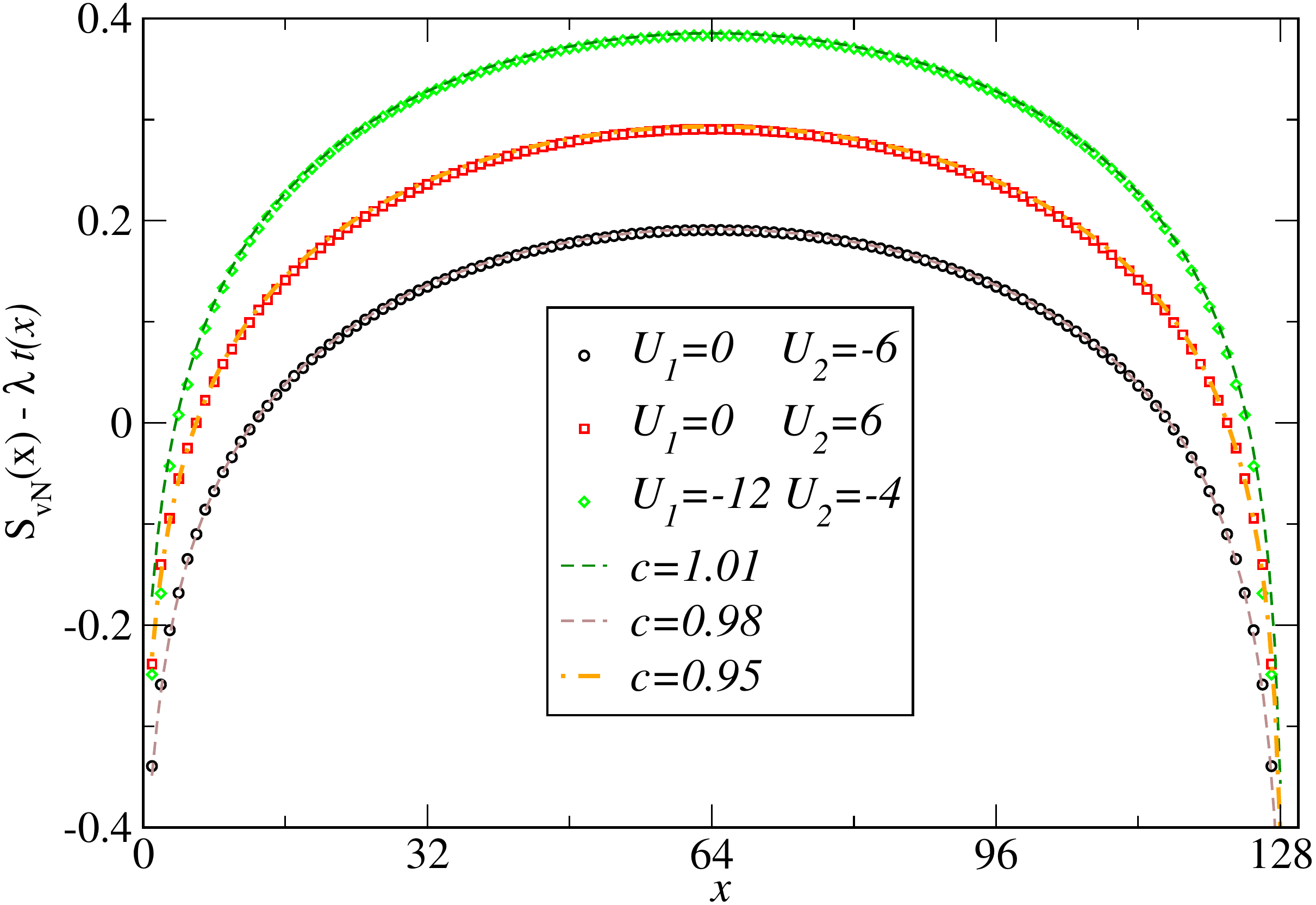}
\includegraphics[width=\columnwidth,clip]{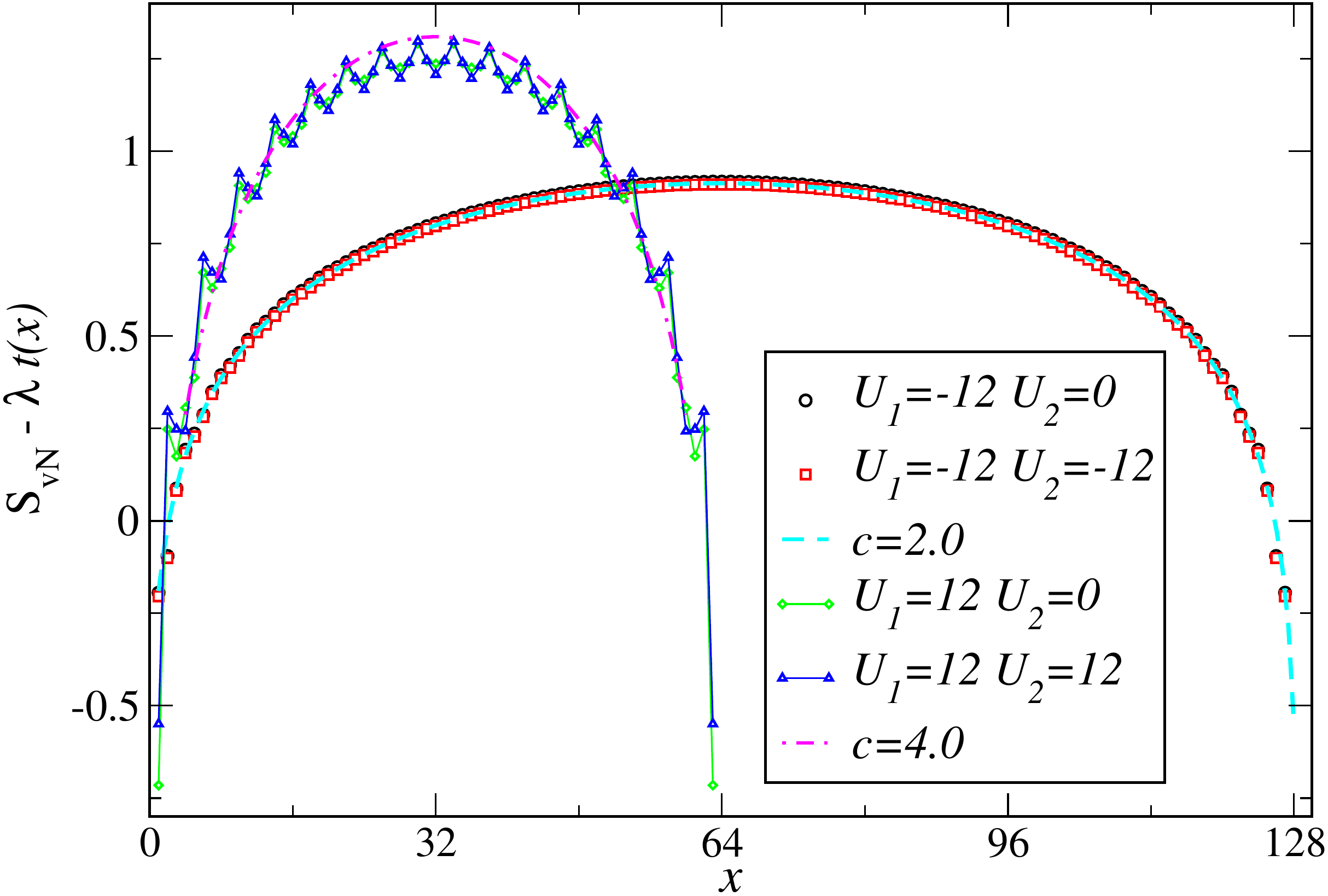}
\caption{(color online)(a) Block entanglement entropies vs size $x$ obtained for $L=128$ and various $(U_1,U_2)$. Oscillations have been removed using the bond kinetic energies $t(x)$ and fitting parameter $\lambda$ (equal respectively to $2$, $1.25$ and $2.4$ (see text). In all cases, we can extract a central charge $c\simeq 1$ as expected. (b) Similar data along the quantum phase transitions which can be fitted with $c=2.0$ (using $\lambda=0.325$) in the attractive case and  with $c=4.0$ (using $\lambda=1.5$ and $1$, plus a shift for clarity) in the repulsive case using smaller $L=64$ system.}
\label{fig:entropies_LL}
\end{figure}

In Fig.~\ref{fig:entropies_LL}(a), we have plotted typical data in phases II, III and IV. By removing oscillations using the bond kinetic energy as an additional fitting parameter, we can obtain smooth functions that perfectly agree with the expected behavior (\ref{eq:Svn}) with $c=1$. 

Now, considering the expected quantum phase transition along the lines $U_1=U_2$ and $U_2=0$, we can clearly identify different behaviors for attractive vs repulsive interactions (as expected since there will be a finite spin gap or not respectively). In Fig.~\ref{fig:entropies_LL}(b) for attractive interactions, we can fit our entanglement entropies data perfectly with $c=2$ using $L=128$. In the repulsive case, our numerical data are not converged on the same size keeping up to $m=8000$ states, so that we plot instead data obtained with $L=64$ and $m=8000$. In this case, we cannot remove entirely the oscillations in the data, but we can get a very good fit using the expected $c=4$ central charge. Overall, we  get an excellent agreement with the theoretical predictions. 

%%%%%%%%%%%%%%%%%%%%%%%%%%%%%%%%%%%%%%%%%%%%%%%%%%%%%%%
\section{Conclusion}\label{sec:conclusion}
%%%%%%%%%%%%%%%%%%%%%%%%%%%%%%%%%%%%%%%%%%%%%%%%%%%%%%%

We have presented a comprehensive study of the most general model relevant for one-dimensional $p$-band two-component fermionic cold gases with local interactions only. We have concentrated here on incommensurate filling and quarter-filling.~\footnote{The half-filled case was already studied in great details in Ref.~\onlinecite{bois}.}

Using a state-of-the-art low-energy approach, supplemented with a one-loop RG numerical analysis, we have found that generically, the charge sector decouples from the spin-orbital one which is gapped. As a consequence, most of the phase diagram is occupied by standard Luttinger liquid phases with a single gapless charge mode (hence a central charge $c=1$). Nevertheless, we have clarified the nature of the dominant instability and have found that it is always of some BCS superconducting kind, in one of the following channels: BCS$_{\mathrm{bond}}$, BCS$_-$, BCS$_+$ and BCS$_d$, see Eqs.~(\ref{BCSs})-(\ref{BCS-})-(\ref{BCS+})-(\ref{BCSd}) respectively. In particular, an interesting bond-centered superconducting instability emerges along the harmonic line for the repulsive interaction. The nature of the phase transitions between these four superconducting phases is also elucidated and found to behave with central charges $c=2$ or $c=4$. 

Our numerical simulations do confirm that phases with dominant BCS$_-$, BCS$_+$ and BCS$_d$ extend from weak to strong coupling and occupy large regions in the phase diagram. In particular, for attractive interactions and harmonic trapping, BCS$_+$ correlations are the dominant ones, different from the $2k_F$-CDW phase predicted earlier~\cite{Kobayashi2014}.
In the last region where the bond-centered superconducting instability BCS$_{\mathrm{bond}}$ is expected at weak coupling, our DMRG data at quarter-filling (i.e., one particle per site) indicate that a Mott phase intervene with fully gapped bond-ordering waves at intermediate coupling, and spontaneous polarization at strong coupling. 
We have also numerically confirmed the nature of all the quantum phase transitions present in this model. 

The $p$-band two-component fermion mode, studied in this paper, 
is thus an interesting model to explore superconducting instabilities of doped two-leg fermionic ladder systems as well as the occurence of a Mott transition. Given the recent progress in realizing SU($N$)-symmetric Fermi gases~\cite{CazalillaRey,review}, we hope that some of the phases discussed here could be realized experimentally and probed using local spectroscopy techniques or by measuring short-range correlations.

%%%%%%%%%%%%%%%%%%%%%%%%%%%%%%%%%%%%%%%%%%%%%%%%%%%%%%%%%%%%%%%%%%%%%%%%%%%%%%%%
\section*{Acknowledgements}
The authors would like to thank K. Totsuka for a related collaboration and useful discussions. 
Numerical simulations have been performed using HPC resources from GENCI--TGCC, GENCI--IDRIS (Grant x2015050225) and CALMIP (grant 2015-P0677). The authors would like to thank CNRS for financial support (PICS grant).

%%%%%%%%%%%%%%%%%%%%%%%%%%%%%%%%%%%%%%%%%%%%%%%%%%%%%%%%%%%%%%%%%%%%%%%%%%%%%%%%

%%%%%%%%% BIB-files %%%%%%%%%%%%%%%%%%%%%%%%%%%%%%%%%%%%%%%%
\bibliographystyle{apsrev4-1}
%\bibliography{./Ref_pbandN2}
\include{paperbandN2.bbl}

%%%%%%%%%%%%%%%%%%%%%%%%%%%%%%%%%%%%%%%%%%%%%%%%%%%%%%
\end{document}